\documentclass{article}

\usepackage{hyperref}
\usepackage{color}
\usepackage{graphicx}
\usepackage{amstext,amsmath,amssymb,amsfonts,bbm,
amsthm}
\usepackage{fancyhdr}
\usepackage{subfigure}
\usepackage{float}

\begin{document}


\title{Light propagating in a Born-Infeld background as seen by an accelerated observer}

\author{Elda Guzman-Herrera\footnote{eguzman@fis.cinvestav.mx} and 
Nora Breton\footnote{nora@fis.cinvestav.mx} \vspace{1cm}\\
Departamento de F\'{i}sica \\
Centro de Investigaci\'{o}n y de Estudios Avanzados del I. P. N. \\ 
Apdo. 14-740, CDMX, M\'{e}xico}

\maketitle

\begin{abstract}
We study the propagation of light in the Born-Infeld (BI) background as seen by an accelerated observer.
In a Born-Infeld electromagnetic field, light trajectories are governed by the null geodesics of the effective optical metric.
The accelerated observer is in a Rindler frame, a situation that, according to the Einstein Equivalence Principle,  is equivalent to being in a uniform gravitational field. 
The phase velocity of light propagating through a purely magnetic or electric BI background is determined as measured by the Rindler observer. The BI field and the acceleration of the frame have opposite effects on the propagating light: while the intense electric or magnetic BI background slows down the velocity of light, in the accelerated frame light may exceed its velocity in vacuum.
We consider light propagating parallel or transversal to the acceleration direction of the Rindler frame. The redshift of light pulses sent from one Rindler observer to another, in the BI background, is calculated as well.
\end{abstract}
\vspace{1cm}

PACS: 12.20.Ds, 11.10.Wx, 41.20.Jb

\newpage

\section{Introduction}
\label{sec:Introduction}

In the presence of intense electromagnetic fields quantum electrodynamics (QED) predicts that vacuum has properties of a material medium as a consequence of the electromagnetic field self-interactions. Then nonlinear electromagnetic (NLEM) effects like light-light interaction or pair production from vacuum excited by an electric field arise.  These effects become significant when the electromagnetic field strengths approach 
$E_{cr} \approx m_e^2 c^3/(e \hbar) \approx 10^{18}$ Volt/m or $B_{cr} \approx 10^{9}$ Tesla; 
$B_{\rm cr}$ represents the field at which the cyclotron energy equals $m_e c^2$, and defines the field scale at which the impact of the external field on quantum processes becomes significant.

These NLEM interactions can be described in an effective way by  Lagrangians ${\mathcal{L}}_{\rm NLED} (F,G) $ that depend nonlinearly on the two Lorentz and gauge invariants of the Faraday tensor $F_{\mu \lambda}$, $F=F^{\mu \lambda}F_{\mu \lambda}= 2(B^2-E^2)$ and $G=F^{*\mu\lambda}F_{\mu\lambda}= -4 \vec{B} \cdot \vec{E}$.
There are several such NLEM proposals, but two of them stand out:  the Euler–Heisenberg (EH) and the Born-Infeld (BI) theories. EH theory was derived from QED principles by W. Heisenberg and H. Euler in 1936 \cite{EH}; for a nice discussion on the history of the Euler-Heisenberg approach see \cite{Dunne2012} and a pedagogical review can be found in \cite{Dunne2004}. 
By treating the vacuum as a medium, EH effective action predicts rates of nonlinear light interaction processes since
it takes into account vacuum polarization to one loop and is valid for electromagnetic fields that change slowly compared to the inverse electron mass. 

The other relevant NLED theory is due by M. Born and L. Infeld. 
Born and Infeld (1934) \cite{BI1934} presented a theory with nonlinear corrections to Maxwell electrodynamics but from a classical perspective.  They intended to contribute to the discussion on the nature of the electromagnetic mass of charged particles; at that time the opinions were divided between if the mass of a charged particle is a manifestation of the electromagnetic field or if field and particle are two separate entities. The aim as well was of mending the singularity of the field and energy of a point charge at its position, proposing then, the existence of a maximum attainable electromagnetic field, given by the BI parameter $b$, with a magnitude  
$b= e / r_0^2 = 10^{20}$ Volt/m, where $r_0$ is the classical electron radius. This is a classical theory that models in an effective way vacuum polarization as a material medium, in this sense resembling EH theory. Another interesting feature is that it does present neither birefringence nor shock waves. The BI Lagrangian is given by

\begin{equation}
{\mathcal{L}}_{\rm BI} (F,G) =    -4b^2 \left\{ 1 - \sqrt{1+ \frac{F}{2b^2}  - \frac{G^2}{16b^4}} \right\}.
\label{BILagrangian}
\end{equation}
The linear electromagnetic Maxwell theory is recovered in the limit that $b \mapsto \infty$, then $\mathcal{L}_{\rm Maxwell} (F) = F$.

Efforts are currently in progress for measuring some of the NLED effects, we mention just a few of them:
Light by light interactions can be studied using heavy-ion collisions; the electromagnetic (EM) field strengths produced, for example by a Pb nucleus would be up to $10^{25}\rm{Vm^{-1}}$, those intense EM fields can be treated as a beam of quasi-real photons, and it has been measured light by light scattering in $Pb+ Pb$ collisions at the Large Hadron Collider \cite{Atlas2017}. Other experimental evidence includes the measurement of photon splitting in strong magnetic fields \cite{Akhmadaliev2002}; the search for vacuum polarization with laser beams crossing magnetic fields or the detection of vacuum birefringence with intense laser pulses \cite{Luiten2004}. There is also the detection of QED vacuum nonlinearities using waveguides \cite{Brodin2001}. Vacuum pair production, known as the Sauter-Schwinger effect \cite{Schwinger1951},  was a prediction in the EH 1936 paper, however, the necessary electric field strengths,  corresponding to a critical laser intensity of about $I_{\rm cr}= 4.3 \times 10^{29}$W$/ cm^2$ \cite{Gelis2016} are not reached yet experimentally.
The situation of strong magnetic fields has an astrophysical interest as well, neutron stars can possess magnetic fields in the range of $10^6-10^9$ Tesla, then processes like photon splitting and pair conversion are expected to occur in their vicinity \cite{Baring2008} \cite{Mignani2017}.

On the other hand, it is well known that an electromagnetic wave traveling through intense EM fields reduces its phase velocity due to vacuum polarization. 
This subject has been addressed since the seventies in the literature \cite{Ritus1972}, \cite{Bialynicka1970}, as well as recently, see for instance \cite{Hu2007}, \cite{Guzman2021} for the EH theory while in \cite{Aiello2007} for the wave propagating in a BI background. 
In \cite{Perlick2015} is proposed the use of a Michelson-Morley interferometer for measuring the changes in the phase velocity due to NLEM theories, in specific the Born-Infeld theory; they concluded that for a BI parameter of the order $b\approx10^{20} {\rm Volt/m}$, the intensity of the background fields in question is still not accessible to near-future experiments, instead, using realistic intensities of background fields ($B\approx 1{\rm T}$), the interferometric experiments could place bounds on the BI parameter $b$ to an order of $10^{14} -10^{15} {\rm Volt/m}$. 

In this paper, we study the propagation of an electromagnetic wave in the BI-NLED background as seen by an accelerated observer. This setup is interesting to investigate because according to the equivalence principle, an accelerated frame is equivalent to a uniform gravitational field, and after all, we are always under the influence of such a gravitational environment. A light ray moving in such an accelerated frame will modify its velocity and pulses will be redshifted. We determine the phase velocity of light, its expression showing the interplay of the magnetic (electric) background and the acceleration of the frame. In the limit of zero acceleration, we recover the wave propagating in the BI background, and in the absence of BI field, we recover the Rindler propagation and the corresponding frequency shifts.
 
The paper is organized as follows: In the next section we
 present the effective optical metric, whose null geodesics are the light trajectories, as well as the phase velocity for a  BI electromagnetic background. In Section 3 we present the Rindler spacetime and the coordinate transformation that connects the Minkowski metric with the accelerated frame. In Section 4 we derive the phase velocity of light propagating through a purely magnetic BI  background from the point of view of the accelerated observer (uniform gravitational field).  In Section 5 we analyze the phase velocity of light propagating in a purely electric BI  background from the point of view of the Rindler observer. In both cases we consider light propagating parallel and perpendicular to the accelerated frame. The limits of zero acceleration and vanishing  BI electromagnetic field are also presented.  Section 6 is devoted to determining the redshift of the propagating light, and how the presence of the BI electromagnetic field affects the frequency shift.
 Finally, the Conclusions are presented in the last Section 7.

\section {Effective optical metric and  phase velocity of light  in a BI background}
\label{sec:EffMetAndPhaseVel}

It is well known that intense EM fields, where the Maxwell theory is no longer valid, can resemble a curved spacetime, in the sense that light trajectories are not straight lines but suffer deflection. Deviations from the straight trajectories in vacuum are described in NLED by the null trajectories of an effective optical metric.  The effective optical metric is obtained from the analysis of the propagation of discontinuities or perturbations of the EM field \cite{Pleban},  \cite{Novello2000},  \cite{Novello2000b}, i. e. according to this formalism the EM fields of the propagating wave are much smaller than the background fields.
The effective optical metric approach turns out to be equivalent to the soft photon approximation.
Splitting the total electromagnetic field into a background field $F_{\mu \nu}$ and a propagating photon $f_{\mu \nu}$, and keeping the linear approximation with respect to $f_{\mu \nu}$ in the equations of motion,  leads to an eigenvalue equation for the propagating
modes  \cite{Hu2007}, \cite{Liberati-Sonego-Visser}.

 Considering that $k_{\mu}$ is a null vector normal to the characteristic surface of the wave, the effective optical metric $g_{\rm eff}^{\mu\nu}$ is given by

\begin{equation}
\label{effgen}
 g_{\rm eff}^{(i) \mu\nu}k_{\mu}k_{\nu}=0, \quad i=1,2,
\end{equation}
where the $(i)$ superscript corresponds to the two metrics that can arise in NLED, when the phenomenon of birefringence occurs. 
The equations of the propagation of the field discontinuities in nonlinear electrodynamics characterized by a Lagrangian $\mathcal{L}(F,G)$  are given in   \cite{Novello2000}, [Eqs. (28)-(29)], and in \cite{Novello2000b} [Eqs. (16)]. Analyzing the propagation of linear waves associated with the discontinuity of the field in the limit of geometrical optics De Lorenci et al. obtain the pair of coupled equations

\begin{equation}
\label{effgen1}
   \zeta k^2=\frac{4}{L_{F}}F^{\lambda \nu}F^{\mu}{}_{\lambda}k_{\nu}k_{\mu}(L_{FF}\zeta + L_{FG}\zeta^{*})-\frac{G}{L_{F}}k^2 (L_{FG}\zeta+L_{GG}\zeta^{*})
\end{equation}

\begin{equation}
\label{effgen2}
    \zeta^{*}k^2=\frac{4}{L_{F}}F^{\lambda\nu}F^{\mu}{}_{\lambda}k_{\nu}k_{\mu}(L_{FG}\zeta+L_{GG}\zeta^{*})-\frac{G}{L_{F}}k^2(L_{FF}\zeta+L_{FG}\zeta^{*})+2\frac{F}{L_{F}}k^2(L_{FG}\zeta+L_{GG}\zeta^{*})
\end{equation}

For a NLED Lagrangian $\mathcal{L}(F,G)$ if we turn off  either the electric or the magnetic field,  $\vec{E}=0$ or $\vec{B}=0$, then $G=4 \vec{E} \cdot \vec{B} =0$, and if additionally the Lagrangian is such that $\mathcal{L}_{FG}=0$, then  the effective metrics (\ref{effgen1},\ref{effgen2}) become 

\begin{equation}
   k^2=\frac{4}{L_{F}}F^{\mu \nu}F^{\tau}{}_{\mu}k_{\nu}k_{\tau}L_{FF} \quad \rightarrow k_{\nu}k_{\mu}\left(L_{F} \eta^{\mu \nu}-4L_{FF}F^{\mu}{}_{\lambda}F^{\lambda \nu} \right)=0
\end{equation}

\begin{equation}
    k^2=\frac{4}{L_{F}}F^{\lambda\nu}F^{\mu}{}_{\lambda}k_{\nu}k_{\mu} L_{GG}+2\frac{F}{L_{F}}k^2 L_{GG}\quad \rightarrow k_{\mu}k_{\nu}\left( ( \mathcal{L}_{F} - 2 \mathcal{L}_{GG} F )\eta^{\mu\nu}-4\mathcal{L}_{GG}F^{\mu}{}_{\lambda}F^{\lambda\nu}\right)=0
\end{equation}
defining the effective metrics for a Lagrangian $\mathcal{L}=\mathcal{L}(F)$ as
\begin{eqnarray}
\label{eff1}
g_{\rm eff}^{(1) \mu\nu} &=& ( \mathcal{L}_{F} - 2 \mathcal{L}_{GG} F )\eta^{\mu\nu}-4\mathcal{L}_{GG}F^{\mu}{}_{\lambda}F^{\lambda\nu}, \\
\label{eff2}
g_{\rm eff}^{(2) \mu\nu} &=& \mathcal{L}_{F}\eta^{\mu\nu}-4\mathcal{L}_{FF}F^{\mu}{}_{\lambda}F^{\lambda\nu},
\end{eqnarray}
where the subscript in the Lagrangian indicates derivative respect to that invariant,
$\mathcal{L}_{X} = d \mathcal{L} / dX$  and $\eta^{\mu \nu} = {\rm diag} [+1,-1,-1,-1]$ is the Minkowski metric. In the Maxwell case  $\mathcal{L}=F,\quad \mathcal{L}_{F}=1, \quad \mathcal{L}_{FF}= 0 $ and $\mathcal{L}_{G}= 0$, then both effective metrics become  conformal to the Minkowski metric, $g_{\rm eff}^{(1) \mu\nu} = g_{\rm eff}^{(2) \mu\nu}= \eta^{ \mu\nu}$, and the null geodesics coincide with the light trajectories in the Minkowski  spacetime.

See \cite{Obukov2002} for a study on the Fresnel equation in nonlinear electrodynamics and \cite{Goulart2009} for a classification of the effective metrics. 
 The Born-Infeld non linear theory is characterized by the non existence of birefringence, this means that the two metrics  in Eqs. (\ref{eff1})-(\ref{eff2}) become conformal, 
$g_{\rm eff}^{(1) \mu\nu}= \Omega^2 g_{\rm eff}^{(2) \mu\nu}$, and we know that a conformal factor does not alter null geodesics. 
Therefore in what follows we shall omit the superscript $(i), \quad i=1,2$. For the BI case, the effective optical metric is given by

\begin{equation}
g_{\rm eff}^{ \mu\nu}= \left(b^2+ \frac{F}{2} \right) \eta^{ \mu\nu}+ F^{\mu}{}_{\lambda}F^{\lambda \nu},    
\label{BI_effmetr}
\end{equation}
where the EM invariant $F=2(B^2-E^2)$.

For a light ray propagating along the $j$-direction, $j=x,y,z$, with wave frequency  
$\omega$ and wave number $k^{j}=k$, its phase velocity can be obtained from 
Eqs. (\ref{effgen}), that is a dispersion relation, as

\begin{eqnarray}
\label{disp_rel}
&& g_{\rm eff}^{ tt} \omega^2 + 2 g_{\rm eff}^{tj} \omega k  + g_{\rm eff}^{ jj} k^2 =0, \\
v_{\rm ph}= \frac{\omega}{k}  & = &  \frac{g_{\rm eff}^{ tj}}{g_{\rm eff}^{ tt}} \pm \sqrt{ \left( \frac{g_{\rm eff}^{ tj}}{g_{\rm eff}^{ tt}}\right)^2 - \left({ \frac{g_{\rm eff}^{ jj}}{g_{\rm eff}^{ tt}}} \right) },  \quad j=x,y,z. 
\label{disp_rel1}
\end{eqnarray}

Since our goal is to study the propagation of light as seen by an accelerated observer, 
for completeness, in the next section, we introduce the accelerated frame or Rindler spacetime and some of its relevant features.
 

\section{The accelerated frame or the Rindler spacetime.}
\label{sec:AcceleratedFrame}

We are interested in the situation of BI electromagnetic fields as seen by observers in a uniform gravitational field, and according to the Einstein Equivalence Principle  (EEP),  a gravitational field can be (locally) modeled by an accelerated frame. 
An example of \textquotedblleft fake\textquotedblright gravity or accelerated frame is the Rindler space,
that we briefly address in this section.

\begin{figure}[H]
\centering
\includegraphics[width=0.6\textwidth]{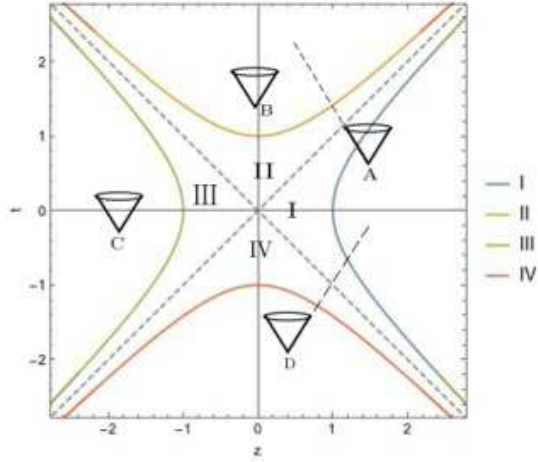}
\caption{\small The dashed diagonal lines represent the null planes at $z=\pm t$ (horizons). An accelerated observer, in region I, follows the blue trajectory; he can receive information from regions I and IV and can send signals to regions I and II. Region III is space-like with respect to the whole Region I, and in particular to the worldline of the observer.}
\label{f:Fig1}
\end{figure}

Let us consider an accelerated observer with constant acceleration $a=\sqrt{a_{\mu}a^{\mu}}$ in $\hat{z}$ direction, and proper time $\tau$; 
the Minkowski  coordinates $(t,z)$ are  related to the accelerated observer  by

\begin{equation}
 t=\frac{1}{a}{\rm Sh}{(a \tau)}; \qquad z=\frac{1}{a}{\rm Ch}{(a \tau)},
 \label{tzRindler1}
\end{equation}
where we denote $\cosh(x)={\rm Ch}(x)$ and $\sinh(x)={\rm Sh}(x)$. Such that 
the line element in  Minkowski  coordinates are related to the Rindler space by \cite{Boulware1980}: 

\begin{equation}
 ds^2=dt^2-dx^2-dy^2-dz^2=(1+a Z)^2dT^2-dZ^2-dx^2-dy^2,.
\label{Rindlermetric}
\end{equation}
 From this expression we see that the coordinate transformation does not cover the entire Minkowski spacetime: at $Z=\frac{-1}{a}$ the metric becomes degenerated; therefore the range of the coordinates must be $Z \in (\frac{-1}{a},\infty)$, $T \in (-\infty, \infty)$, and these coordinates cover only a part of the Minkowski spacetime, denoted as Region I  in Fig.\ref{f:Fig1};  the different regions are separated by event horizons located at $z=\pm t$. 

 Eqs. (\ref{tzRindler1}) represent a hyperbolic curve in Minkowski spacetime,  with the semimajor axis at $1/a$, 

\begin{equation}
z^2-t^2=\frac{1}{a^2}.   
\end{equation}
therefore the accelerated observer´s trajectory is a hyperbola as shown in Fig \ref{f:Fig1}: it starts at $z \rightarrow \infty$, then slows down as it approaches the origin, and at a finite distance from the origin it returns and speeds up as it approaches $z\rightarrow \infty$  \cite{Gupta1998}. Light rays are shown traveling along   $45^{o}$  null lines.   The observer can \textquotedblleft send signals\textquotedblright  towards the upper quadrant  II, as represented by the cone $A$, but the observer in quadrant I cannot receive signals from the upper quadrant II. The observer in the Rindler frame cannot access any signal beyond the horizon, located at $z = \pm t$,   but can receive signals from the inferior quadrant  IV, as illustrated with the light cone $D$, however, A cannot send signals towards D  \cite{Boulware1980, Almeida2006}. 

To determine what does the accelerated observer see, we have to make a coordinate transformation from the Minkowski frame, that we can call as well the Laboratory frame or Lab frame,  to the accelerated observer's frame.  In the accelerated frame we shall describe the phenomena from the point of view of a privileged  observer at $Z=0$  with proper time $\tau =T$, but the general transformation from Minkowski to a Rindler observer located at $Z_{i}$ with acceleration $a_{i}$ is given  by \cite{Kooks2020,Formiga2007},

\begin{equation}
  t_{i}=(\frac{1}{a_{i}}+Z_{i}){\rm Sh} ({a_{i} T_{i}}); \qquad z_{i}=(\frac{1}{a_{i}}+Z_{i}){\rm Ch}({a_{i} T_{i}}).
\end{equation}
 
\begin{figure}[H]
 \centering
 \subfigure{\includegraphics[width=0.55\textwidth]{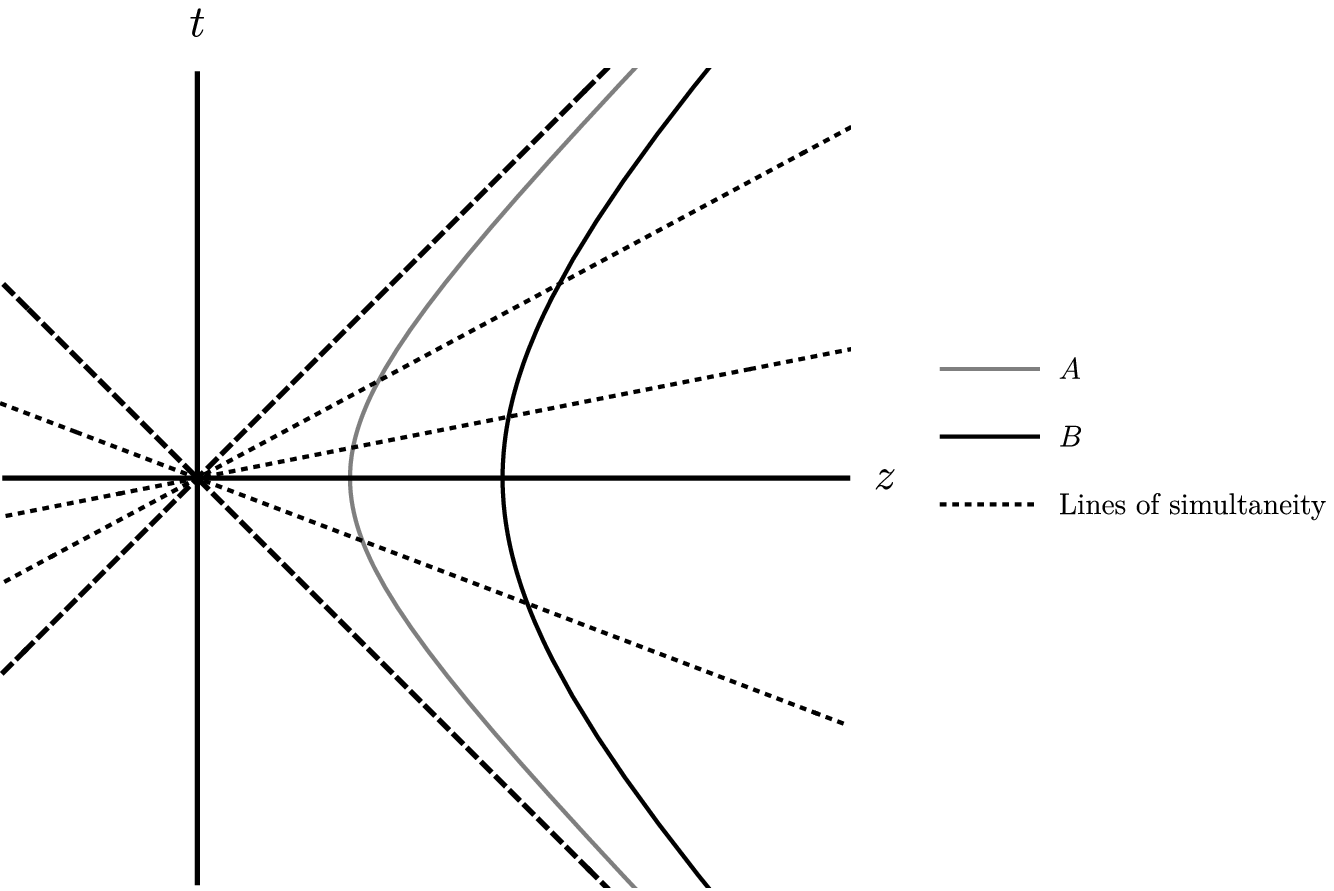}}
 \subfigure{\includegraphics[width=0.35\textwidth]{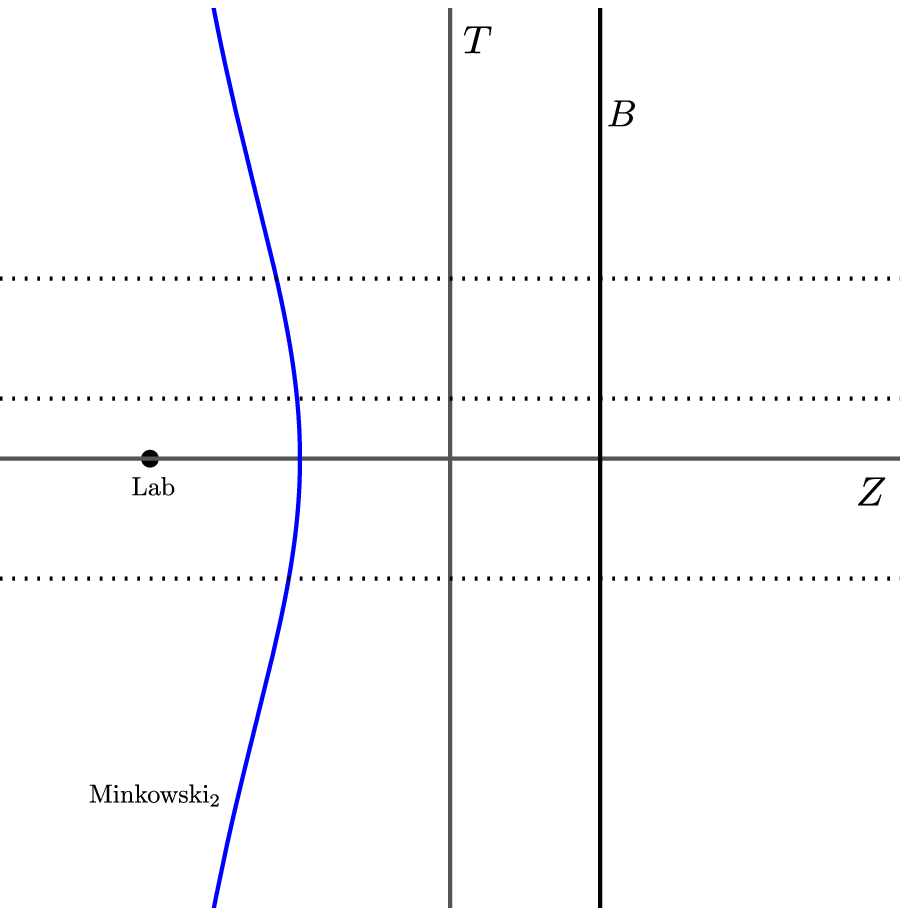}}
\caption{\small The worldlines of two observers $A$ and $B$ in the Minkowski space are shown as the gray and black thick lines in the figure to the left. The dotted lines represent the lines of simultaneity in the Rindler frame. To the left are illustrated the observers in the Minkowski space-time, and to the right are the trajectories in the Rindler frame. We are considering $A$ as the principal observer,  its trajectory is a vertical straight line (fixed $Z$); the dot at $\{Z=-\frac{1}{a},T=0 \}$ is the Lab frame;  the observer $B$ is illustrated in the black thick line; the blue curve corresponds to a third Rindler observer, located between $Z=0$ and $Z=1/a_{A}$. }
\label{f:Fig2}
\end{figure}

Note that taking $Z=0$ and $T=\tau$  Eqs. (\ref{tzRindler1}) are recovered.
Other Rindler observers (other hyperbolas)  have different accelerations $a_{i}$, and different $z_{i}$ ( Lab coordinates) as well.
In the Rindler frame, they have different $Z_{i}$,  but they all have the same velocity when their world lines intersect a line of simultaneity, as is shown in Fig. \ref{f:Fig2}, where the lines of simultaneity are the dotted green lines.  
The proper position of each observer is $Z_{i}= 0$ and $\{t= {\rm Sh} (a_{i}T)/a_i, {\rm Ch} (a_{i}T)/a_i  \}$; then in the Lab frame each observer has a different proper acceleration. Then to say "Uniformly accelerated reference frame",  referring to the Rindler space, does not imply that all the observers have the same acceleration;  in the Lab frame, these observers do not form a rigid lattice, as they do in the Rindler frame.



To define simultaneity, we consider that for each uniformly accelerated frame there is a momentarily co-movil inertial frame so that, along the line of simultaneity, both frames have the 4-velocity parallel to each other, and they measure each other at rest, then it is said that they form a rigid lattice of observers who all agree on simultaneity, \cite{Kooks2020}.  
 In Rindler spacetime the lines that define the simultaneity are $T=n=\text{constant}$ (shown in Fig. \ref{f:Fig2}), and correspond to   $t=z {\rm Th}{(a n) }$ in Minkowski spacetime. 

Let $\tau_{A}$ be the proper time of the principal observer, it dictates what the other observer's clocks measure. When $\tau_{A}=0$ all the observers coincide with the $z$ axis: $T=0,\quad t=0$. Let us distinguish the proper time $\tau$ and the coordinate time $T$. The Rindler line element is given by

\begin{equation}
 ds^2=g_{TT}dT^2+g_{i j}dx^i dx^j , \qquad i,j=\{X,Y,Z \}
\end{equation}

The differential proper time $ d \tau$ and  the spatial line element,$dl^2= g_{ij}dx^{i} dx^{j}$, are such that
$dl^2=-d\tau^2\mid_{dT=0}$ and the coordinate velocity of a Rindler observer is 

\begin{equation}
 v^2=\frac{-g_{ij} dx^i dx^j}{dT^2},
 \label{vline}
\end{equation}
$v^2$ is non negative since we are taken the Minkowski metric as $\eta_{ \mu  \nu} = {\rm diag} [+1, -1, -1, -1 ]$.

For the  null trajectories,  $d\tau^2=0$, 

\begin{equation}
\label{null_geod}
 d\tau^2  =  g_{T T }dT^2 +g_{ij} dx^i dx^j=0,
\end{equation}
and the phase velocity for a light ray traveling in the $z$-direction  in the Rindler frame  is

\begin{equation}
v_{\rm ph}= \frac{dZ}{dT}=\sqrt{-\frac{g_{TT}}{g_{ZZ}}}  = (1+a Z),
 \label{velocityRindler}
\end{equation}
and we can interpret $d\tau/dT$ as the rate of flow of time \cite{Kooks2020} in the Rindler space, then the speed of light equals the rate of flow of time.  The velocity now depends on the position of the observer; it is zero at the horizon $Z=-\frac{1}{a}$ and when $Z=0$ the velocity is the one in vacuum, $c=1$;  the position $Z=0$  corresponds to the ``principal observer". From Eq. (\ref{velocityRindler}) we see that for an observer at 
$Z>0$, he measures a light velocity greater than $c=1$. This result is explained using the global line of simultaneity which is rotating in the laboratory frame, like a radar; when it intersects the worldlines it goes faster for an observer at higher $z$ (Lab frame) than for an observer closer to the origin. The interpretation is that the time flows faster for higher $z$ because the proper time of the observers at greater $z$ is aging faster than for observers closer to the origin. Then,  light covers more distance per unit of time as $z$ is greater. However, each accelerated observer measures a local speed of light being $c=1$, therefore any of them could be chosen as the ``master"  or "principal" observer. 
 
 We can visualize the trajectory of light in Rindler spacetime integrating (\ref{velocityRindler}).

\begin{equation}
 Z=\frac{1}{a}\left[-1+(1+a Z_{0})\exp{a\left( T-T_{0}\right)}  \right],
 \label{TrajNullR}
\end{equation}
when $Z\rightarrow-\frac{1}{a}$, the velocity tends to zero, as expected.  Then, the measurement of  light velocity must be done locally, i.e. at $Z=0$, the position of the ``master" observer.

In the following two sections, \ref{sec:MagneticCase}
 and \ref{sec:ElectricCase}, we address the propagation of light in purely magnetic and purely electric BI backgrounds, respectively, as seen in the Rindler frame.


\section{Light propagating through a Born-Infeld magnetic background in the Rindler space. }
\label{sec:MagneticCase}

In this section, we determine the phase velocity of light rays propagating in a purely magnetic BI background as seen by an accelerated observer.
According to the Einstein Equivalence Principle (EEP), an accelerated frame is equivalent to a uniform gravitational field,  therefore our treatment describes the propagation of light under the influence of a very intense  BI electric or magnetic background, as seen by an observer in a uniform gravitational field.

Light trajectories through an intense magnetic field are the null geodesics of  the effective optical  metric $g_{\rm eff}^{\mu\nu}$, Eq.  (\ref{BI_effmetr}). 
We write down the effective optical metric corresponding to a BI  uniform magnetic field $B$ and then transform it into the Rindler accelerated frame. 

Considering that
the nonvanishing electromagnetic tensor components of the magnetic background 
are $F^{xy}=-B_z, \quad F^{xz}= B_y,\quad F^{yz}=-B_x,$ and $F=2 B^2$, $g_{\rm eff}^{\mu\nu}$ is given by

\begin{equation}
g_{\rm eff}^{\mu\nu}= 
\left(
\begin{array}{cccc}
 b^2+B^2 & 0 & 0 & 0 \\
 0 & -b^2-B_{x}^2 & -B_{x} B_{y} & -B_{x} B_{z} \\
 0 & -B_{x} B_{y} & -b^2-B_{y}^2 & -B_{y} B_{z} \\
 0 & -B_{x} B_{z} & -B_{y} B_{z} & -b^2-B_{z}^2 \\
\end{array}
\right).
\label{MeffEst}
\end{equation}
We shall consider that
the uniform magnetic background is taken, with no loss of generality, in the XZ plane as $\vec{B}= B_{x} \hat{x} + B_{z} \hat{z}= B \sin \theta \hat{x} + B \cos \theta \hat{z}$.  From Eq. (\ref{disp_rel1}),  the phase velocity  $v_{\rm ph}$  of light propagating through a BI magnetic background, along the $\hat{z}$-direction,   with wave frequency $\omega$ and wave number $k^{z}=k$, amounts to

\begin{eqnarray}
v_{\rm ph}^2 &=& 1 - \frac{B_{x}^2}{b^2+B^2}, 
\label{velAiello}
\end{eqnarray}
where  $B^2= B_x^2+B_z^2$.  Then the effect of the BI magnetic background is of slowing down the phase velocity, unless the magnetic component transversal to the propagating light is zero, in which case the phase velocity is the one of light in vacuum. This case was studied in \cite{Aiello2007}.

To determine the phase velocity measured by a Rindler observer the effective optical metric Eq. (\ref{MeffEst}) is transformed to the Rindler frame.  To transform to a Rindler frame with acceleration $\vec{a}= a \hat{z}$ the transformation matrix is given by

\begin{equation}
 R^{\mu}{}_{\alpha}= \frac{\partial x^{\mu}}{\partial X^{\alpha}}=
  \left(
\begin{array}{cccc}
 (1+a Z) {\rm Ch} (a T) & 0 & 0 & {\rm Sh} (a T) \\
 0 & 1 & 0 & 0 \\
 0 & 0 & 1 & 0 \\
 (1+a Z) {\rm Sh} (a T) & 0 & 0 & {\rm Ch} (a T) \\
\end{array}
\right).
\label{TRMatrix}
\end{equation}

Then the transformed effective optical metric, obtained from  $ g_{\rm eff,R}^{\mu\nu}= R^{\mu}_{{}\alpha}  g_{\rm eff}^{\alpha \beta} R_{\beta}^{{} \nu}$ is 

\begin{equation}
 g_{\rm eff,R}^{\mu\nu}=
 \left(
\begin{array}{cccc}
 \frac{b^2+B_{z}^2+\left(B_{\perp}^2\right) {\rm Ch} ^2(a T)}{(a Z+1)^2} & \frac{B_{x} B_{z} {\rm Sh}(a T)}{a Z+1} & \frac{B_{y} B_{z} {\rm Sh} (a T)}{a Z+1} &-\frac{\left(B_{\perp}^2\right) { \rm Sh} (2 a T)}{2(1+a Z)}\\
 \frac{B_{x} B_{z} { \rm Sh} (a T)}{a Z+1} & -\left(b^2+B_{x}^2\right) & -B_{x} B_{y} & -B_{x} B_{z} {\rm Ch} (a T) \\
 \frac{B_{y} B_{z} {\rm Sh} (a T)}{a Z+1} & -B_{x} B_{y} & -\left(b^2+B_{y}^2\right) & -B_{y} B_{z} {\rm Ch} (a T) \\
 -\frac{\left(B_{\perp}^2\right) {\rm Sh} (2 a T)}{2(1+a Z)} & -B_{x} B_{z} {\rm Ch}(a T) & -B_{y} B_{z} {\rm Ch} (a T) & B_{\perp}^2 {\rm Sh} ^2(a T)-b^2-B_{z}^2 \\
\end{array}
\right),
\end{equation}
where the superscript  $``R"$ denotes the effective optical metric  $ g_{\rm eff}^{\mu\nu}$  in the  Rindler frame and $B_{\perp}^2=B_{x}^2+B_{y}^2$. In case $a=0$ the effective optical metric Eq. (\ref{MeffEst}) is recovered.
As a consequence of the transformation to the magnetic background, electric components arise, in a similar way than from a Lorentz boost. The nonvanishing electromagnetic tensor components are now $F_{R}^{tx}=\frac{B_{y} {\rm Sh} (a T)}{1+a Z}$, $F_{R}^{ty}=-\frac{B_{x} {\rm Sh} (a T)}{1+a Z}$, $F_{R}^{xz}=B_{y} {\rm Ch} (a T)$, $F_{R}^{yz}=-B_{x} {\rm Ch} (a T)$, $F_{R}^{xy}=-B_{z}$, and the invariants are preserved, $F=2B^2$, $G=0$. 

The corresponding covariant effective optical metric is 

\begin{equation}
 g^{\rm eff,R}_{\mu\nu}=
 \left(
\begin{array}{cccc}
 (a Z+1) \left[b^2-B_{\perp}^2 {\rm Sh} ^2(a T)\right] & B_{x} B_{z} {\rm Sh} (a T) & B_{y} B_{z} {\rm Sh} (a T) & -\frac{1}{2} B_{\perp}^2 {\rm Sh} (2 a T) \\
 B_{x} B_{z} {\rm Sh} (a T) & -\frac{b^2+B_{y}^2+B_{z}^2}{a Z+1} & \frac{B_{x} B_{y}}{a Z+1} & \frac{B_{x} B_{z} {\rm Ch} (a T)}{a Z+1} \\
 B_{y} B_{z} {\rm Sh} (a T) & \frac{B_{x} B_{y}}{a Z+1} & -\frac{b^2+B_{x}^2+B_{z}^2}{a Z+1} & \frac{B_{y} B_{z} {\rm Ch} (a T)}{a Z+1} \\
 -\frac{1}{2} B_{\perp}^2 {\rm Sh} (2 a T) & \frac{B_{x} B_{z} {\rm Ch} (a T)}{a Z+1} & \frac{B_{y} B_{z} {\rm Ch} (a T)}{a Z+1} & -\frac{B_{\perp}^2 {\rm Ch} ^2(a T)+b^2}{a Z+1} \\
\end{array}
\right)
\end{equation}

\begin{figure}[H]
 \centering
 \includegraphics[width=0.4\textwidth]{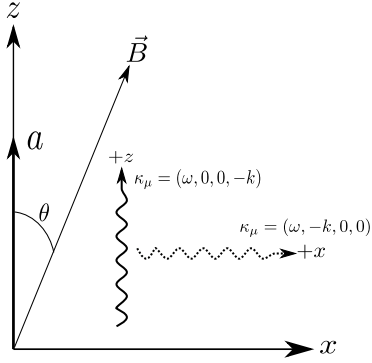}
 \caption{\small Scheme of the relative directions: the proper acceleration of the Rindler frame is $a$ in the $\hat{z}$ direction;  the wavy lines indicate two propagating waves,
  one  (thick) traveling  in $\hat{z}$-direction and another (dotted) in $\hat{x}$-direction. The magnetic field makes a $\theta$ angle with the acceleration direction. }
 \label{f:Fig3}
\end{figure}

For  an EM wave in the $i-$direction with wave number  $\kappa^{\mu}_{R}=(\omega_{R},k_{R}^{i})$,  the phase velocity is given by Eq.  (\ref{disp_rel1});   the acceleration of the Rindler frame is $\vec{a}= a \hat{z}$  and we consider successively  three different propagating directions of the wave:   $+ \hat{z}$ , $- \hat{z}$  and  $+ \hat{x}$; the setting is shown in Fig. \ref{f:Fig3}.
We analyze the corresponding phase velocities in the next subsections.

\subsection{Magnetic  BI background. Light propagating in the $\pm \hat{z}$ direction}
\label{subsec:zP}

For a wave traveling in the $\pm \hat{z}$ direction: $\kappa_{\mu}=\left(\omega,0,0,\mp  k \right)$
the phase velocity is,  from Eq.  (\ref{disp_rel1}),  given by,

\begin{equation}
\frac{v^{\rm R \pm \hat{z}}_{\rm ph}}{1+a Z}= \frac{\omega}{k}= \frac{\mp  B_{x}^2 {\rm Sh} ( a T) {\rm Ch} (aT)  +\sqrt{ \left(b^2+B^2\right) \left(b^2+B_{z}^2\right)}}{ \left(b^2+B^2+B_{x}^2 {\rm Sh} ^2(a T)\right)},
\label{vphi_from_eff_metrRT}
\end{equation}
where the $\mp$    sign corresponds to the wave propagating in   $+ \hat{z}$ and to the wave in $ - \hat{z}$ direction, respectively, and $B^2=B_{x}^2+B_{z}^2$. 
From this expression we can see that $v^{\rm R}_{\rm ph}$ depends on $aT$ and $aZ$, i.e. the effect of increasing time is equivalent to increasing the acceleration of the frame; recall that the Rindler observer will remain at a fixed $Z$. Also, the acceleration affects the phase velocity only if there is a magnetic component that is transversal to the acceleration  (in this case $B_x$). 
In the limit, $a \mapsto 0$ the expression for the pure BI magnetic field Eq. (\ref{velAiello}) is recovered.
Note that the only difference between the wave propagating along  $+ \hat{z}$  and  $ - \hat{z}$ is the sign in the first term in Eq. (\ref{vphi_from_eff_metrRT}),  and it indicates that Rindler frame distinguishes light direction. Moreover, it implies that the phase velocity for the wave along  $ - \hat{z}$ will be greater than the one traveling in the opposite direction. 

Expanding $v^{\rm R}_{\rm ph}$  Eq. (\ref{vphi_from_eff_metrRT}), in powers of $B^2/b^2$ and  keeping  terms of order $\mathcal{O}(1/b^2)$, i.e. neglecting terms of order $\mathcal{O}(1/b^4)$ and higher, we arrive at

\begin{equation}
\frac{v^{\rm R \pm \hat{z}}_{\rm ph}}{1+a Z}=  
1- \frac{B_{x}^2}{2 b^2} \left\{1 - 2 {\rm Sh} (aT) \left[  {\rm Sh} (aT) \pm  {\rm Ch}(aT) \right] \right\};
\label{vphi_from_eff_metrRTapprox}
\end{equation} 
this approximation is still valid for very strong fields; if we consider, for instance, that $B^2/b^2 \approx 10^{-2}$, then the BI field $B$ is of the order $10^{19}$V/m,  that is ten times the critical Schwinger field  $B_{\rm cr} \approx 10^{9}$ Tesla. 
In the limit, $a \mapsto 0$ the expression for the pure BI magnetic field Eq. (\ref{velAiello}) is recovered, up to terms in 
$\mathcal{O} (1/b^2)$.  Notice that even if $B \mapsto b$, the phase velocity reaches a minimum of half the velocity of light in vacuum. The linear limit (no BI field) is obtained with $b \mapsto \infty$ and then $ (v^R_{\rm{ph}})^2=(1+aZ)^2$ is recovered. 

\begin{figure}[H]
 \centering
 \subfigure{\includegraphics[width=0.45\textwidth]{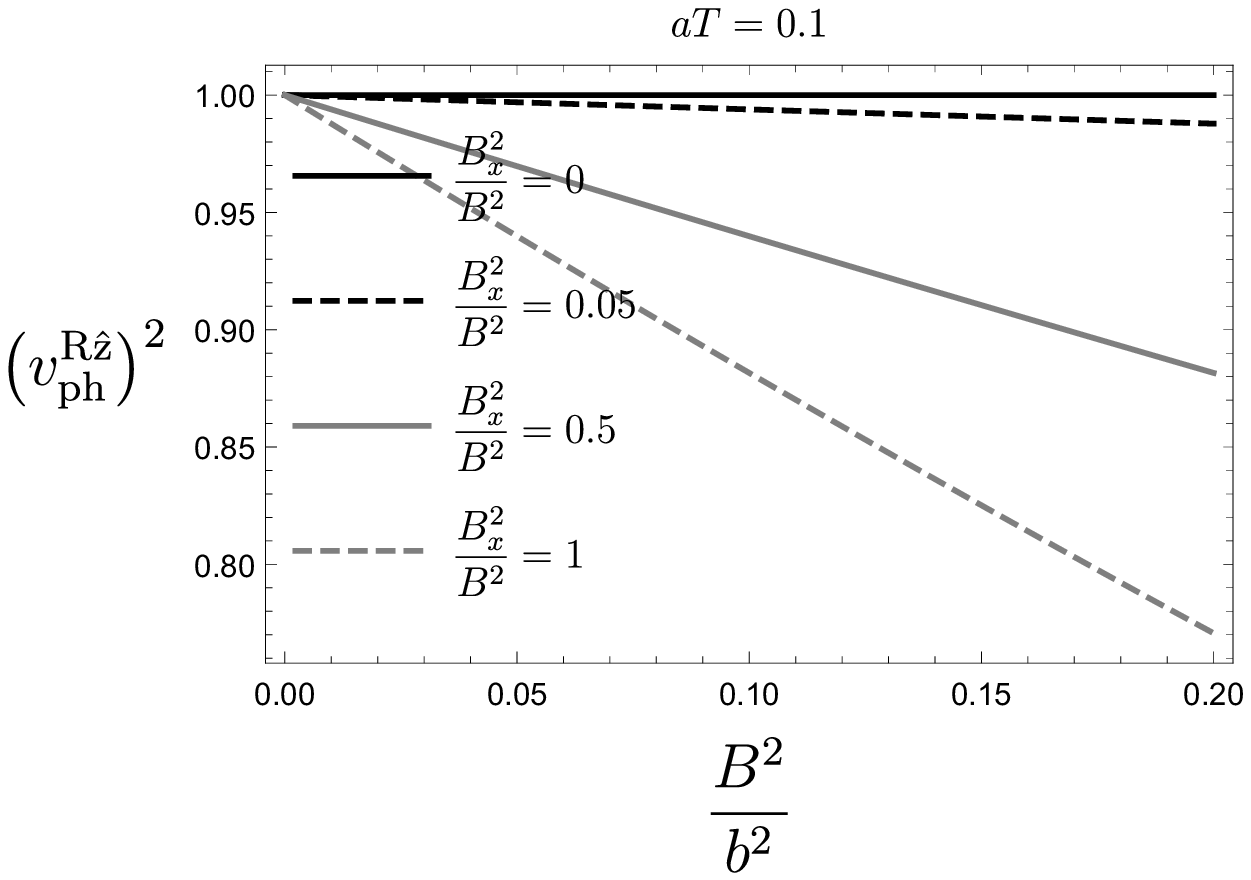}}
 \subfigure{\includegraphics[width=0.45\textwidth]{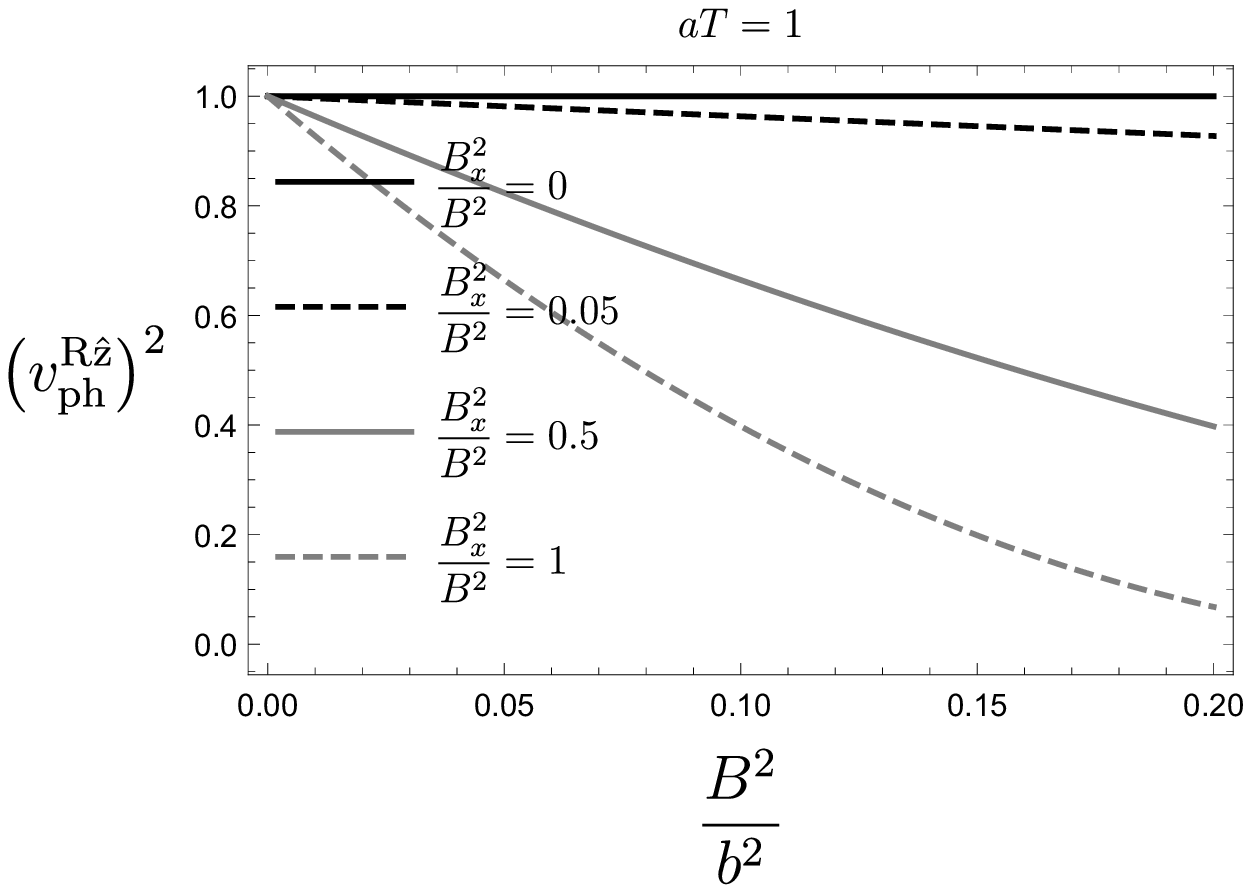}}
 \caption{\small The plots  for the  squared phase velocity of a light ray propagating in $z$  direction through a BI magnetic background as seen by an observer with acceleration $a$ in the $\hat{z}$  direction;  the plots are for different values of $aT$; as $aT$ increases,  the velocity slows down at a faster pace.  The intensities of the perpendicular magnetic components  are varied as shown and $B^2=B_{x}^2+ B_{z}^2$.}
 \label{f:Fig4}
\end{figure}

\begin{figure}[H]
 \centering
 \subfigure{\includegraphics[width=0.45\textwidth]{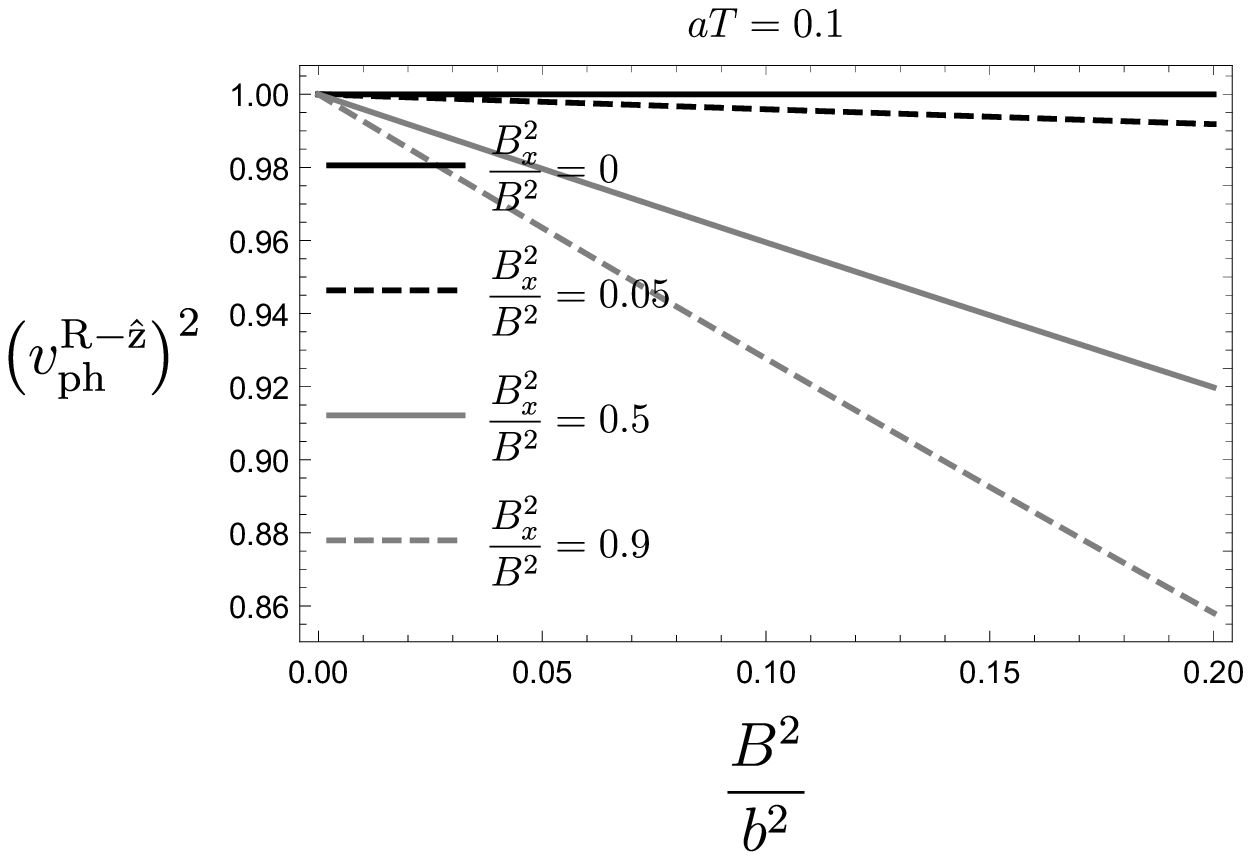}}
 \subfigure{\includegraphics[width=0.45\textwidth]{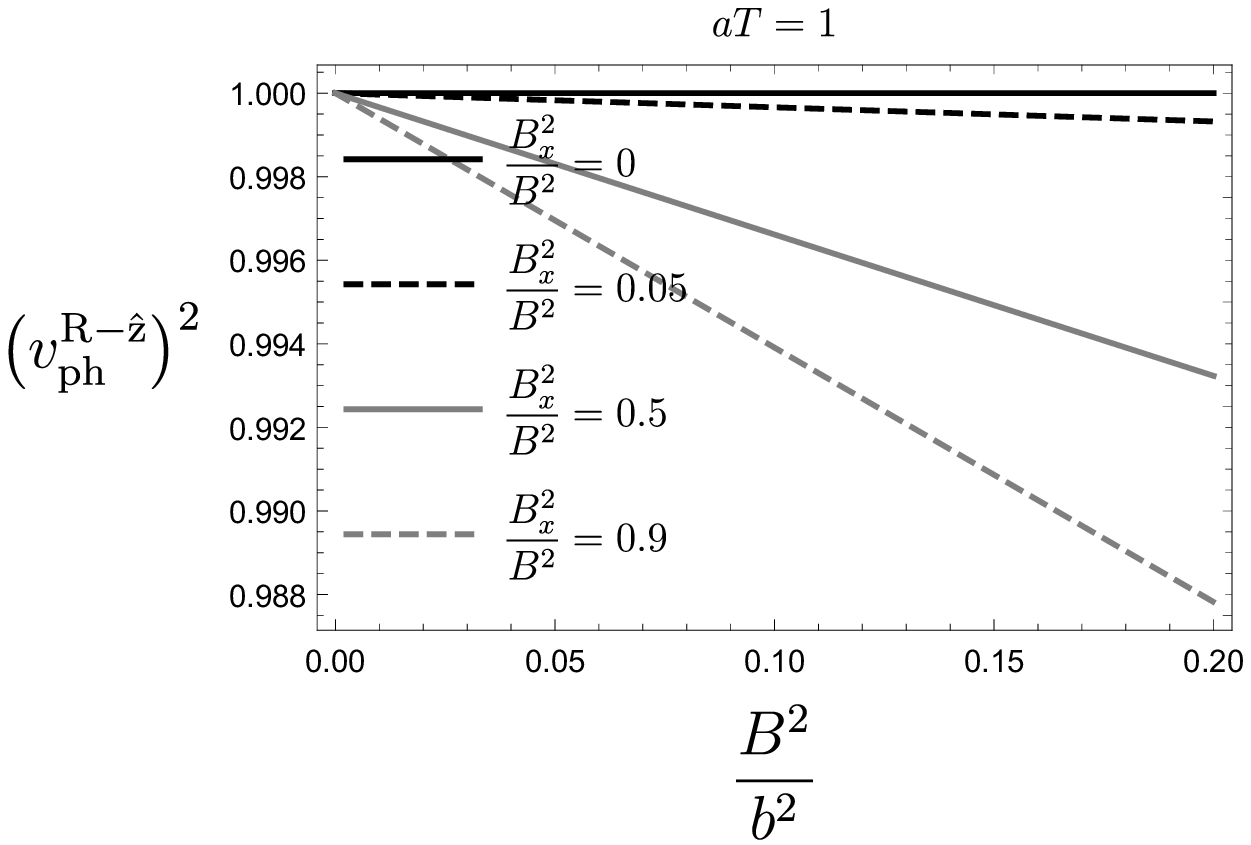}}
 \caption{\small  Phase velocities for a wave moving in the $- \hat{z}$ direction for different values of $aT$;  $v_{ph}$ is smaller as the magnetic field  increases.   As $aT$ increases,  the velocity slows down at a slower pace.  The intensities of the perpendicular magnetic components  are varied as shown and $B^2=B_{x}^2+ B_{z}^2$. }
 \label{f:Fig5}
\end{figure}

Considering in Eq. (\ref{vphi_from_eff_metrRTapprox}) that   the term that depends on $aT$, for the wave traveling in $+ \hat{z}$,  increases monotonically   (${\rm Sh} (aT) \left[  {\rm Sh} (aT) +  {\rm Ch}(aT) \right]$; while  for the wave traveling in $- \hat{z}$, the corresponding term, ${\rm Sh} (aT) \left[  {\rm Sh} (aT) - {\rm Ch}(aT) \right]$,  decreases monotonically up to $-1/2$;  then the effect of the accelerating frame in one case $+ \hat{z}$) contributes to slowing down the wave and in the other  ($- \hat{z}$) it increases the phase velocity. In the last case the Rindler effect is opposite to the BI effect.
Therefore for a Rindler observer accelerated in  $+\hat{z}$ direction the phase velocity of the wave propagating in the $+ \hat{z}$ direction is smaller than the one of the waves along $-\hat{z}$.  That intuitively corresponds to traveling in the same direction of the wave in the first case while in the second case going to the encounter of the wave propagating along  $-\hat{z}$. This effect is superposed with the BI slowing down of the wave and the result is illustrated in Figs. \ref{f:Fig4} and  \ref{f:Fig5}:
as the intensity of the background field increases the light velocity diminishes;
by turning off the BI background field it is recovered the light velocity in vacuum. 
 
Contrary to what happens with the wave propagating in $+ \hat{z}$  direction,
for the wave propagating in $- \hat{z}$  direction the pace of slowing down decreases as $aT$ increases, and in the limit of long times or large acceleration, $aT \mapsto \infty$, the BI effect of slowing down is lost and the Rindler phase velocity $ (v_{\rm ph}^{R})^2=(1+aZ)^2$ is recovered.


\subsection{Magnetic BI background. Light propagating in the $+ \hat{x}$ direction}
\label{subsec:x}

For a wave moving in the $\hat{x}$ direction with wave number $\kappa^{R}_{\mu}=\left(\omega_{R},-k_{R},0,0 \right)$,   the magnetic background component along $\hat{x}$, $B_{x}$, is parallel to the propagating ray and perpendicular to the Rindler acceleration, while $B_{z}$ is perpendicular to the light direction.   The phase velocity,  from Eq. (\ref{disp_rel1}), is given by 

\begin{equation}
\frac{v^{\rm R  \hat{x}}_{\rm ph}}{\left(1+a Z\right)} = \frac{B_{x} B_{z} {\rm Sh} (a T)  +\sqrt{\left(b^2+B^2\right) \left(b^2 + B_{x}^2 {\rm Ch}^2(a T)\right)}}{b^2+B^2+B_{x}^2 {\rm Sh}^2(a T)}.
\label{v_ph_x}
\end{equation}
In this case the effect of turning off the $B_{x}$ component  eliminates the terms depending on  $aT$, being this the manifestation
that the only significative magnetic field is the one transversal to the acceleration of the frame.

\begin{figure}[H]
 \centering
 \subfigure{\includegraphics[width=0.45\textwidth]{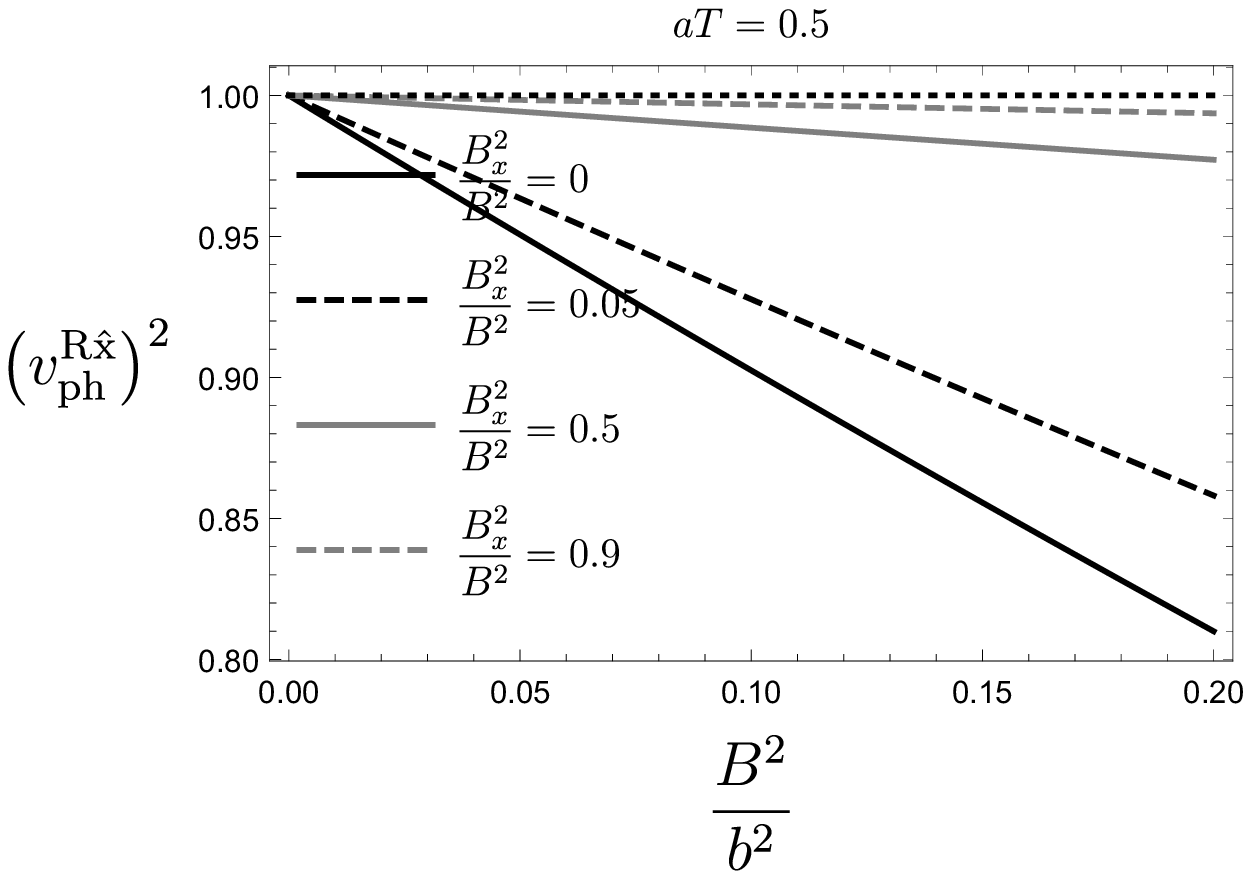}}
 \subfigure{\includegraphics[width=0.45\textwidth]{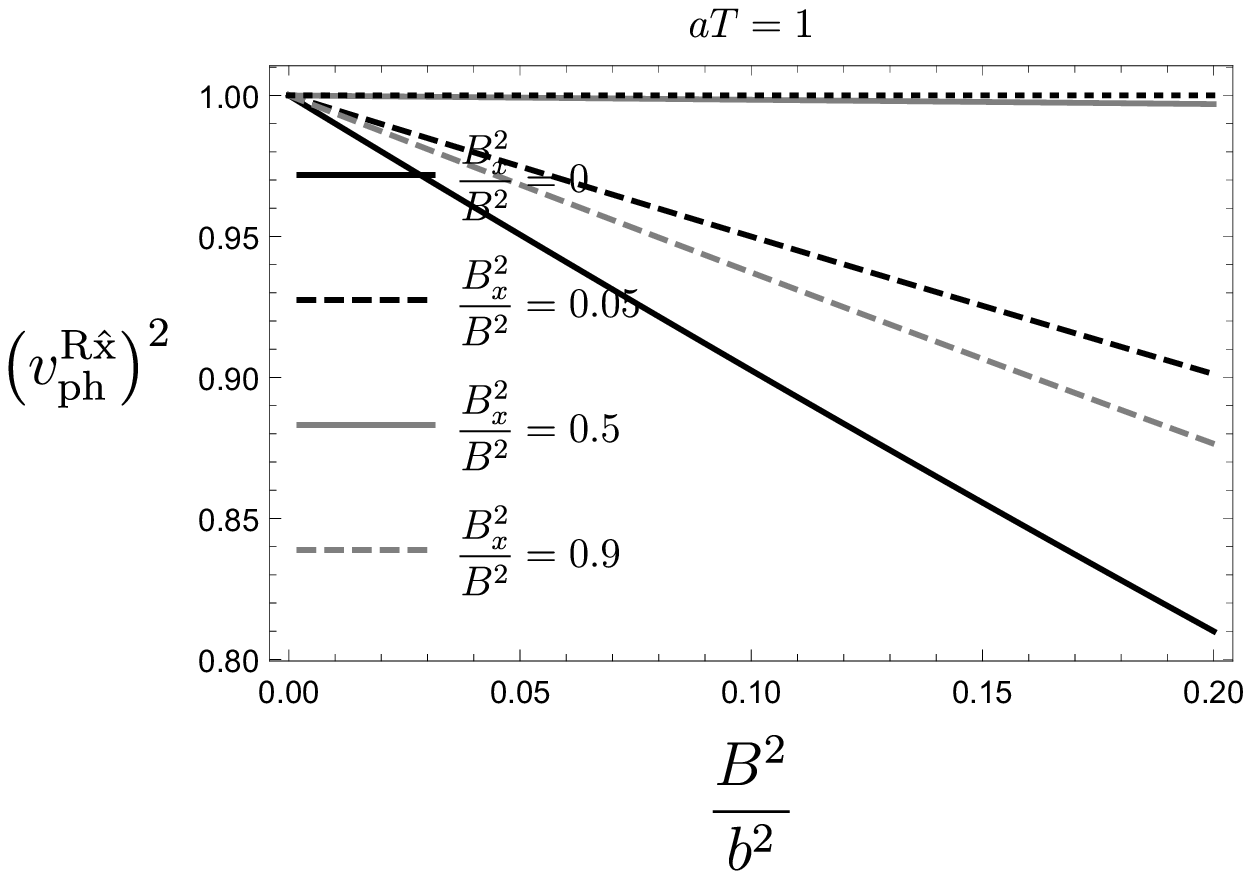}}
 \caption{\small Phase velocities for the wave moving in the $\hat{x}$ direction, for different values of  $aT$. In the plot to the left, for $aT=0.5$, as $B_{x}$ increases the phase velocity  tends to the velocity of light in vacuum. To the right, for $aT=1$,  for large field components, $(B_{x}/B)^2=0.9$ the phase velocity does not approach the one in vacuum.}
 \label{f:Fig6}
\end{figure}

 Making $a=0$ we get
\begin{equation}
v^{\rm R  \hat{x}}_{\rm ph} = \sqrt{1-  \frac{ B_{z}^2}{b^2+B^2}},  
\label{v_ph_x2}
\end{equation}
that corresponds to the pure BI effect of slowing down $v_{\rm ph}$.
Expanding $v^{\rm R \hat{x}}_{\rm ph}$ in powers of $B^2/b^2$ and  keeping  terms of order $\mathcal{O}(1/b^2)$, i.e. neglecting terms of order $\mathcal{O}(1/b^4)$ and higher, we arrive at

\begin{equation}
\frac{v^{\rm R  \hat{x}}_{\rm ph}}{\left(1+a Z\right)} = 1- \frac{1}{2 b^2} \left[ B_{z} - B_{x} {\rm Sh}(a T) \right]^2, 
\label{v_ph_x_approx}
\end{equation}
The linear limit, $b \mapsto \infty$,  gives the Rindler velocity, $(v^{\rm R \hat{x}}_{\rm ph}{})^2 =(1+ a Z)^2$; in this case the acceleration of the frame gives an effect opposite to the BI slowing down but only for a certain range of $(aT)$;  the minimum value of (\ref{v_ph_x_approx}) occurs  for

\begin{equation}
 aT= {\rm ArcSinh} \frac{B_{z}}{B_{x}},
\end{equation}
for $aT$ larger than the one of the maximum,  $aT >  {\rm ArcSh} (\frac{B_{z}}{B_{x}} )$, $v^{R \hat{x}}_{\rm ph}$  decreases such that
in the limit of large $aT$ $v^{R \hat{x}}_{\rm ph}$ approaches zero. 
Making zero the component parallel to the acceleration, $B_{z}=0$, the phase velocity renders

\begin{equation}
\frac{v^{\rm R  \hat{x}}_{\rm ph}}{\left(1+a Z\right)} = 1- 
\frac{B_{x}^2}{2b^2}{\rm Sh}^2 (a T);
\end{equation}
while if the magnetic component that is perpendicular to the acceleration vanishes, $B_{x}=0$, 
\begin{equation}
\frac{v^{\rm R  \hat{x}}_{\rm ph}}{\left(1+a Z\right)} = 1- 
\frac{B_{z}^2}{2b^2}.
\end{equation}
Then if we aimed to diminish the most the phase velocity,  the most effective way would be to turn off the component along the Rindler acceleration, $B_{z}=0$,  since $ {\rm Sh(aT)} \ge 0, \quad \forall aT.$
In Fig. \ref{f:Fig6} is plotted the phase velocity of the wave traveling in $\hat{x}$ direction for different values of $aT$. 

\begin{figure}[H]
\centering
\includegraphics[width=0.8\textwidth]{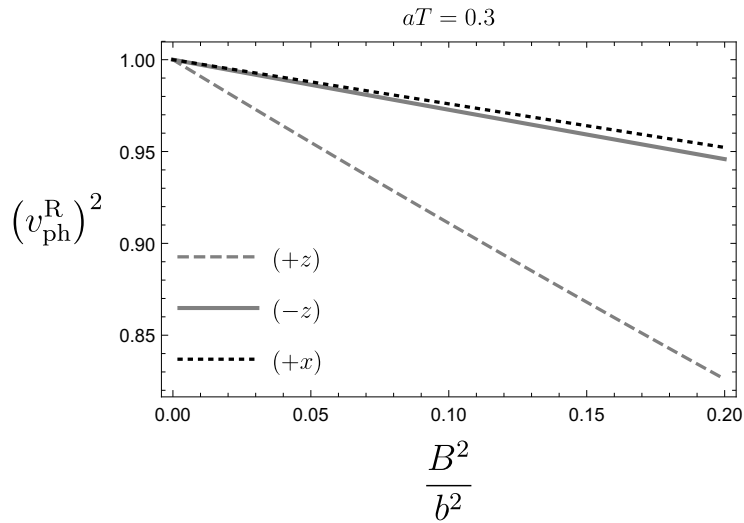}
\caption{\small   The phase velocities for  the three directions of propagation of the light wave in the Rindler frame are shown. The Rindler acceleration is $a \hat{z}$.  The smallest velocity is for the wave moving in the $+ \hat{z}$ direction; for the three directions  $v_{\rm ph}^{R}$ decreases as $B/b$ increases and $aT=0.3$. In this plot the parallel and the transversal magnetic field components have the same intensity $B_{x}=B_{z}=B/\sqrt{2}$.  }
\label{f:Fig7}
\end{figure}

In Fig. \ref{f:Fig7} we compare the phase velocities for the three different directions of the light rays. The smallest velocity is for the wave moving in the $+ \hat{z}$ direction; for the three directions of the light ray $v_{\rm ph}^{R}$ decreases as $B$ increases,  at fixed $(aT)$.

\begin{figure}[H]
\centering
\includegraphics[width=0.7\textwidth]{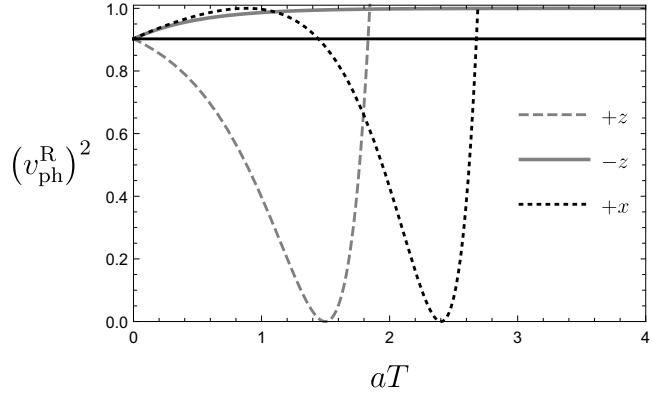}
\caption{\small The squared modulus of phase velocities as a function of $aT$ are shown, for a fixed magnetic background,  $(B/b)^2=0.2$; the magnetic field components are $B_{z} = B_{x} = B /\sqrt{2}$. The black thick line represents the phase velocity of the light ray in presence of the BI magnetic background when $aT=0$.  For the waves moving in the  $+ \hat{z}$ and $+ \hat{x}$ directions   $v^{\rm R}_{\rm ph}$'s decrease up to a minimum and then increase reaching the velocity in vacuum. This corresponds to the Rindler observer's view, which initially is behind the light ray,  reaches it, and then moves away from it; since the accelerated frame is no longer inertial, there is no problem in surpassing light velocity in vacuum.
For light in $- \hat{z}$ direction it reaches $c=1$ increasing monotonically.}
\label{f:Fig8}
\end{figure}

In Fig. \ref{f:Fig8} are shown the phase velocities of the light rays propagating along the three considered directions as a function of $aT$, for fixed $(B/b)^2=0.2$. We are taking the parallel and the transversal magnetic field components with the same intensities $B_{x}=B_{z}=B/ \sqrt{2}$. 

\begin{figure}[H]
\centering
\includegraphics[width=0.5\textwidth]{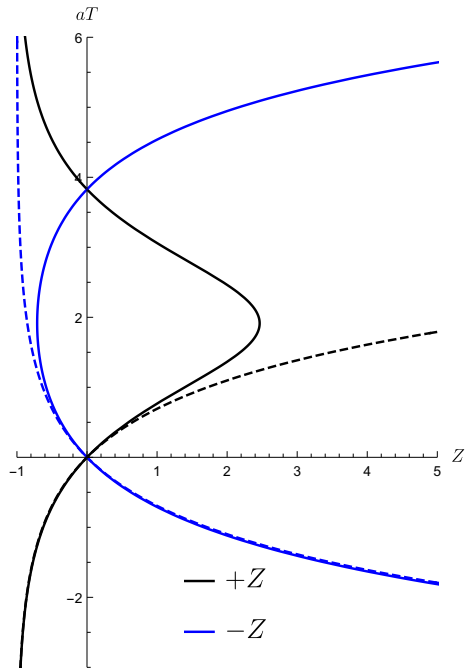}
\caption{\small  The trajectories followed by the $\pm z$ waves are illustrated. The dashed lines represent the trajectories for the waves when there is no BI effect affecting them and the thick lines are the trajectories of the waves in presence of a magnetic background with $B_{x}=B_{z}=B/\sqrt{2}$ and $(B/ b)^2 = 0.2$. }
\label{f:Fig9}
\end{figure}

The behavior of increasing or decreasing depends on the values of $aT$;  in the first range of $aT$,
the velocity is smaller for the wave moving along  $+ \hat{z}$ and closer to $c=1$ for the waves traveling in the  $- \hat{z}$ and  $+ \hat{x}$ directions.
While in a second range of $aT > 2$, $v_{\rm ph}^{R}$ is smaller for the wave moving along  $+ \hat{x}$ and approaches $c=1$ for the waves traveling in the  $\pm \hat{z}$ directions.
 As  $aT$ increases the BI effect of slowing down the phase velocity is canceled for the waves traveling in the  $\pm \hat{z}$ directions.
Recall that these are the velocities as measured by the accelerated observer; this observer moves with respect to the light ray, first to reach it, and later the observer moves away from the light ray. This is possible since the frame is no longer an inertial one. The trajectories of light rays in $ \pm \hat{z}$ directions are shown in Fig. \ref{f:Fig9}.
At the base of the previous analysis we can summarize the behavior of the phase velocity of the light ray propagating through the BI magnetic field as seen by an observer with acceleration $\vec{a}=a \hat{z}$:
The phase velocity depends on $aT$ such that the effect of increasing the acceleration is the same as the one of time elapsing.
 For the waves moving in the  $+ \hat{z}$ and $+ \hat{x}$ directions   $v^{\rm R}_{\rm ph}$'s decrease up to a minimum and then increase reaching the velocity in vacuum, while for light in $- \hat{z}$ direction it reaches $c=1$ departing from the Rindler $v^{\rm R}_{\rm ph}$ and increasing monotonically. The change in direction of light can be understood by thinking in the relative motion of the accelerated observer that initially is behind the light ray, then it reaches it when the accelerated observer measures $v_{\rm ph}^{R}=0$ and then moves away from the light ray (the accelerated observer sees that light ray changed its direction).
If the magnetic field is in the same direction as the Rindler acceleration, then there is no effect on the phase velocity.
For a fixed BI magnetic field, the smallest velocity is for the wave moving in the $+ \hat{z}$ direction. In Figs. \ref{f:Fig7} and \ref{f:Fig8} can be observed these behaviors.

Regarding the question that if the values of the BI field  are reachable in experiments see \cite{Perlick2015}; it is not clear if it would be possible to build such an accelerated frame, or if it would be possible to maintain the acceleration enough time to observe the predicted changes in the phase velocities.

\section{Light propagating through a BI electric background in the Rindler space.}
\label{sec:ElectricCase}

The propagation of an EM wave through an intense uniform electric field
is of interest  since there is the prediction of vacuum electron-positron production \cite{Adler1970}
that has not yet been measured, however, it might be feasible in the near future, due to the high power reached lately by lasers  \cite{Gies2009}, \cite{Karbstein2020}.
Considering that the nonvanishing electromagnetic tensor components of a purely uniform electric background are $F^{xt}=E_{x}$, $F^{yt}=E_{y}$, $F^{zt}=E_{z}$, and the electromagnetic invariant $F=-2E^2$ ($B=0$), the expression for the effective optical metric from Eq. (\ref{BI_effmetr}) is 

\begin{equation}
 g_{\rm eff}^{\mu\nu}=\left(
\begin{array}{cccc}
 b^2 & 0 & 0 & 0 \\
 0 & -b^2+E_{y}^2+E_{z}^2 & -E_{x} E_{y} & -E_{x} E_{z} \\
 0 & -E_{x} E_{y} & -b^2+E_{x}^2+E_{z}^2 & -E_{y} E_{z} \\
 0 & -E_{x} E_{z} & -E_{y} E_{z} & -b^2+E_{\perp}^2 \\
\end{array}
\right).
\end{equation}
 From Eq. (\ref{disp_rel1})  it is determined  the phase velocity of light propagating through a uniform BI electric field as, 

\begin{equation}
 v_{\rm ph}^{2}=1-\frac{E_{\perp}^{2}}{b^2}
 \label{velAielloE}
\end{equation}
with $E_{\perp}^2=E_{x}^2+E_{y}^2$, being  the electric field component  perpendicular to the acceleration direction $\hat{z}$. Note that in absence of the perpendicular component the phase velocity is the one in vacuum. This expression was also derived in \cite{Aiello2007}.
 According to Eq. (\ref{velAielloE}) in principle, the wave could reach zero velocity; however taking the average in polarization and the electric field components,  the minimum accessible phase velocity turns out to be $ <v_{\rm ph}>=1/3$ \cite{Obukov2002}.

Now transforming the effective metric for the electric BI background  to the Rindler frame with (\ref{TRMatrix}), we obtain:
\begin{equation}
 g_{\rm eff,R}^{\mu\nu}=
 \left(
\begin{array}{cccc}
 \frac{E_{\perp}^2 {\rm Sh} ^2(a T)+b^2}{(a Z+1)^2} & \frac{E_{x} E_{z} {\rm Sh} (a T)}{a Z+1} & \frac{E_{y} E_{z} {\rm Sh} (a T)}{a Z+1} & -\frac{E_{\perp}^2 {\rm Sh} (a T) {\rm Ch} (a T)}{a Z+1} \\
 \frac{E_{x} E_{z} {\rm Sh} (a T)}{a Z+1} & -b^2+E_{y}^2+E_{z}^2 & -E_{x} E_{y} & -E_{x} E_{z} {\rm Ch} (a T) \\
 \frac{E_{y} E_{z} {\rm Sh} (a T)}{a Z+1} & -E_{x} E_{y} & -b^2+E_{x}^2+E_{z}^2 & -E_{y} E_{z} {\rm Ch} (a T) \\
 -\frac{E_{\perp}^2 {\rm Sh} (a T) {\rm Ch} (a T)}{a Z+1} & -E_{x} E_{z} {\rm Ch} (a T) & -E_{y} E_{z} {\rm Ch} (a T) & E_{\perp}^2 {\rm Ch}^2(a T)-b^2 \\
\end{array}\right)
\end{equation}
We consider as previously that the frame acceleration is  $\vec{a}=a \hat{z}$ and that, with no loss of generality,   the uniform BI field is located in the XZ plane, + $\vec{E}= E_{x} \hat{x} +E_{z} \hat{z}$; then the phase velocity of the wave moving in the $\pm \hat{z}$ and $+ \hat{x}$ directions is obtained from  Eq. (\ref{disp_rel1})
and analyzed in the next two subsections.
\subsection{Electric BI background. Light propagating in the $ \pm \hat{z}$ direction}
\label{subsec:EzP}

For the wave moving in the $\pm \hat{z}$ direction, and  considering the position of the principal observer at $Z=0$, the phase velocity, resulting from  Eq. (\ref{disp_rel1}), is given by  

\begin{equation}
v^{\rm R \pm \hat{z}}_{\rm ph} = \frac{ \mp E_{x}^2 {\rm Sh}(a T) {\rm Ch} (a T)  +  \sqrt{ b^2 \left(b^2-E_{x}^2\right)}}{b^2+E_{x}^2 {\rm Sh} ^2(a T)}, 
 \label{vRindEzp}
\end{equation}
We are taken the square root with the plus sign to recover the phase velocity in absence of acceleration.
Then the difference in the phase velocity between the wave propagating in  $+ \hat{z}$ direction and the one along $- \hat{z}$ is only the sign in the first term.
The consequence of this difference is that the magnitude of the phase velocity for the wave in the  $+ \hat{z}$ direction is smaller than the one traveling in the opposite direction.
There are several features shared with the magnetic case:  $v^{\rm R}_{\rm ph}$ depends on $aT$ and $aZ$, i.e. the effect of increasing time $T$ or spatial coordinate $Z$  is equivalent to increasing the acceleration of the frame. Also, the acceleration affects the phase velocity only if there is an electric component that is transversal to the acceleration  (in this case $ E_x$); i.e. the acceleration of the frame notices only the electric component that is transversal to $\vec{a}= a \hat{z}$. 
Expanding $v^{\rm R}_{\rm ph}$ in powers of $E^2/b^2$ and  keeping  terms of order $\mathcal{O}(1/b^2)$, i.e. neglecting terms of order $\mathcal{O}(1/b^4)$ and higher, we arrive at

\begin{equation}
 v^{\rm R \pm \hat{z}}_{\rm ph}= 1- \frac{ E_{x}^2}{2 b^2} \left\{ 1+2 {\rm Sh}(a T) \left[ {\rm Sh} (a T) \pm {\rm Ch}(a T) \right] \right\}, 
 \label{vRindEzp_ap}
\end{equation}
comparing the previous equation with the corresponding to the magnetic BI background, Eq. (\ref{vphi_from_eff_metrRTapprox}), we notice that the expression is the same changing the magnetic to the electric component, $B_{x} \mapsto E_{x}$; this means that to that order the application of a magnetic uniform background produces the same effect than the application of an electric one, as long as the transversal  magnetic and electric components have the same magnitude.
Consequently,   the limiting cases are very similar to the ones for the magnetic background: in the limit, $a \mapsto 0$ the expression for the pure BI electric field Eq. (\ref{velAielloE}) is recovered. The linear limit (no BI field) is obtained with $b \mapsto \infty$ and then $ v_{\rm{RT}}^2= 1$ is recovered ($Z=0$).  In the limit of long times or large acceleration, $aT \mapsto \infty$ the BI effect of slowing down the phase velocity is lost, and the Rindler phase velocity $ v_{\rm{RT}}^2=(1+aZ)^2$ is recovered.
In this case, we omit the figures for $v_{\rm ph}^{\rm R}$ for the wave propagating along $\pm \hat{z}$ to not be redundant.


\subsection{Electric BI background. Light propagating in the $+ \hat{x}$ direction}
\label{subsec:Ex}

For the wave moving in the $+ \hat{x}$ direction, the phase velocity is

\begin{equation}
 v^{\rm R \hat{x}}_{\rm ph}=\frac{E_{x} E_{z} {\rm Sh}(a T)  + \sqrt{b^2 \left(E_{x}^2{\rm Sh} ^2(a T)+b^2-E_{z}^2\right)} }{E_{x}^2 {\rm Sh}^2(a T)+b^2}
\label{vRindExp}
\end{equation}
Recall that $E_{x}$ is perpendicular  with respect to the acceleration of the Rindler frame, $\vec{a}=+a\hat{z}$. In this case both components of the BI electric field, the parallel and the perpendicular to the acceleration play a role.  In the case that $E_{x} =0$ there is no effect of the acceleration of the frame, and the wave slows down with the pure BI effect, 
\begin{equation}
 v^{\rm R \hat{x}}_{\rm ph}= \sqrt{ 1- \frac{ E_{z}^2 }{b^2}} \approx  1- \frac{ E_{z}^2 }{2 b^2},
\end{equation}
note that even if $E_{z}=b$ the light velocity does not vanish but there is a lower bound of $v_{\rm ph min}^{R \hat{x}} =1/2$.
Expanding $v^{\rm R \hat{x}}_{\rm ph}$ in powers of $E^2/b^2$ and  keeping  terms of order $\mathcal{O}(1/b^2)$, i.e. neglecting terms of order $\mathcal{O}(1/b^4)$ and higher, we arrive at

\begin{equation}
  v^{\rm R \hat{x}}_{\rm ph} =  1- \frac{1}{2 b^2} \left[ E_{z} - E_{x} {\rm Sh}(a T) \right]^2,
\end{equation}

Note that it is the same expression of   Eq. (\ref{v_ph_x_approx}) with $B_{i} \rightarrow E_{i}$, then as in the $\pm \hat{z}$ directions, the limiting cases are very similar to the ones for the magnetic background, then again we omit the figures for the wave propagating along $\hat{x}$ and the comparison between the considered directions. 

For the light propagating along $+ \hat{z}$ direction  $v^{R}_{\rm ph}$ diminishes monotonically in an interval of $(aT)$, until reaching a minimum that is arbitrarily close to zero, then for $(aT)$ larger than a certain $aT_{c}$ given by

\begin{equation}
   \text{aT}_{c} = \frac{1}{2} {\rm ArcSh} \left(2 \sqrt{\frac{b^2-E_x^2}{E_x^4}}\right), 
   \label{aTE}
\end{equation}
$v_{\rm ph}^{R}$ changes its  propagation direction and   increases reaching   light velocity in vacuum.  For the wave in $+ \hat{x}$ direction, the behavior is qualitatively similar than the one for the wave along $+ \hat{z}$: as  $aT$ increases the wave slows down monotonically until reaching a minimum that is arbitrarily close to zero, then  increases till reaching $v_{\rm ph}^{R}=1$; for $E_{z}=0$ the slowing down is maximized.  For the wave moving in the $- \hat{z}$ direction as $aT$ increases the  phase velocity reaches  the one in vacuum. 
Recall that these are the velocities as measured by the accelerated observer,  and the relative directions change as $aT$ increases:  initially, the observer chases the light ray, and as it approaches the wave, the wave's velocity seems to decrease, and eventually, it is zero (the moment the observer reaches the wave) and then the relative direction changes, since subsequently, the observer moves away from the wave, then wave's velocity starts to increase; in the plot is shown the modulus squared of the phase velocity. Since the accelerated frame is no longer inertial, no special relativity velocity invariance is expected.

If the electric field component that is perpendicular to the acceleration vanishes, then there is no effect neither of the acceleration or the BI field on the phase velocity, and its value is the one in vacuum. 


\subsection{The phase velocities for strong fields}

The series expansion of the phase velocities up to terms of order $B^2/b^2$ gives the same behavior for BI magnetic and electric background. However, it is worth (briefly) discussing the behavior of the phase velocities considering the exact expressions, Eqs. (\ref{vRindEzp}) and (\ref{vRindExp}),
since for strong fields, i.e. when the fields approach the maximum attainable electromagnetic field $b$ differences arise and it can be distinguished between the electric and the magnetic background.

\begin{figure}[H]
 \centering
\subfigure[]{ \includegraphics[width=0.45\textwidth]{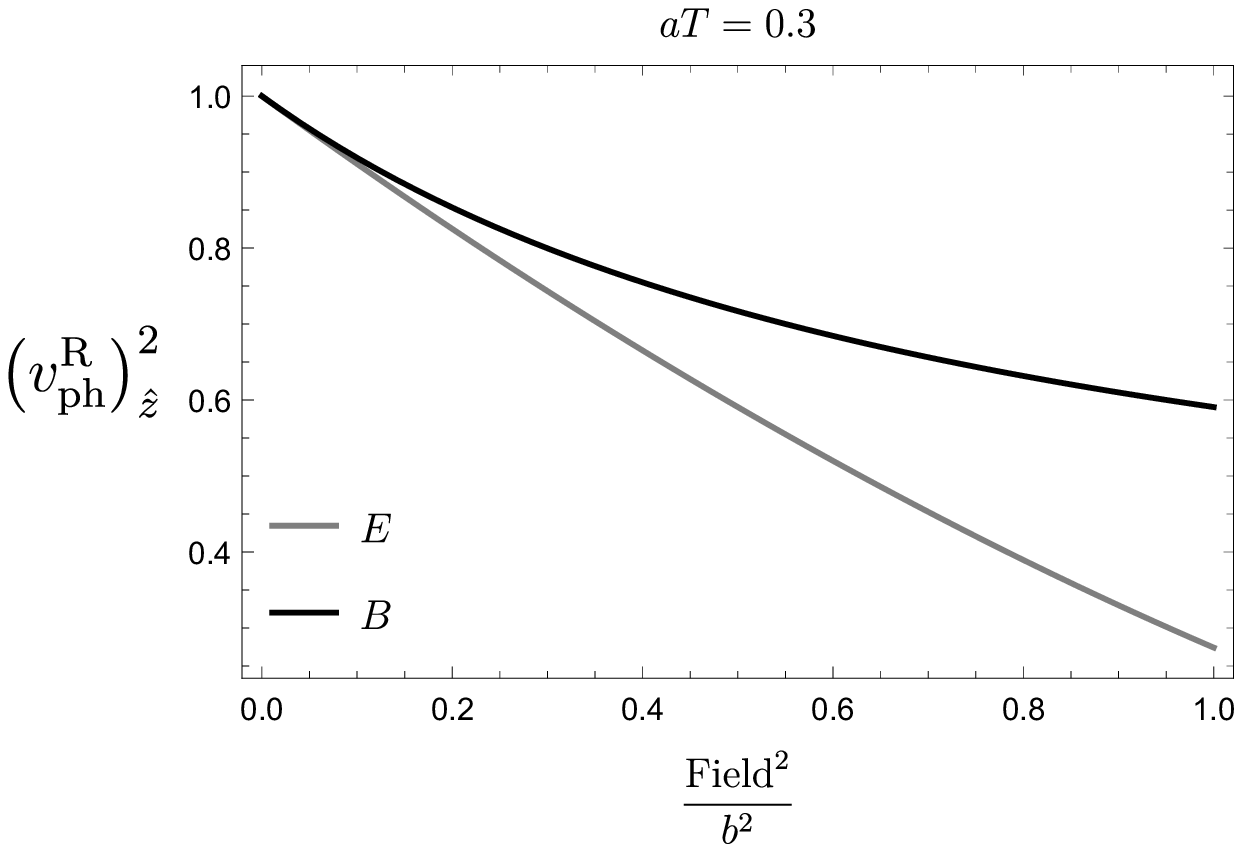}}
\subfigure[]{ \includegraphics[width=0.45\textwidth]{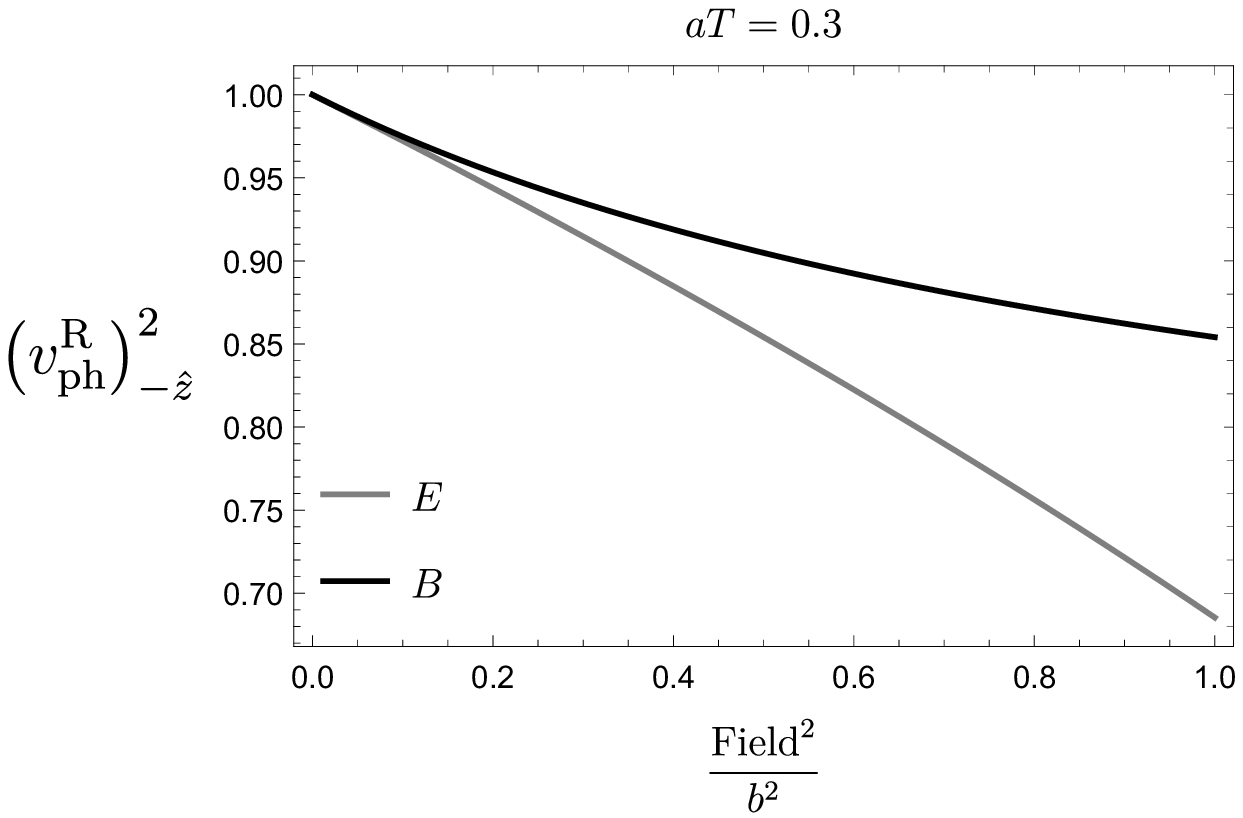}}
\subfigure[]{\includegraphics[width=0.45\textwidth]{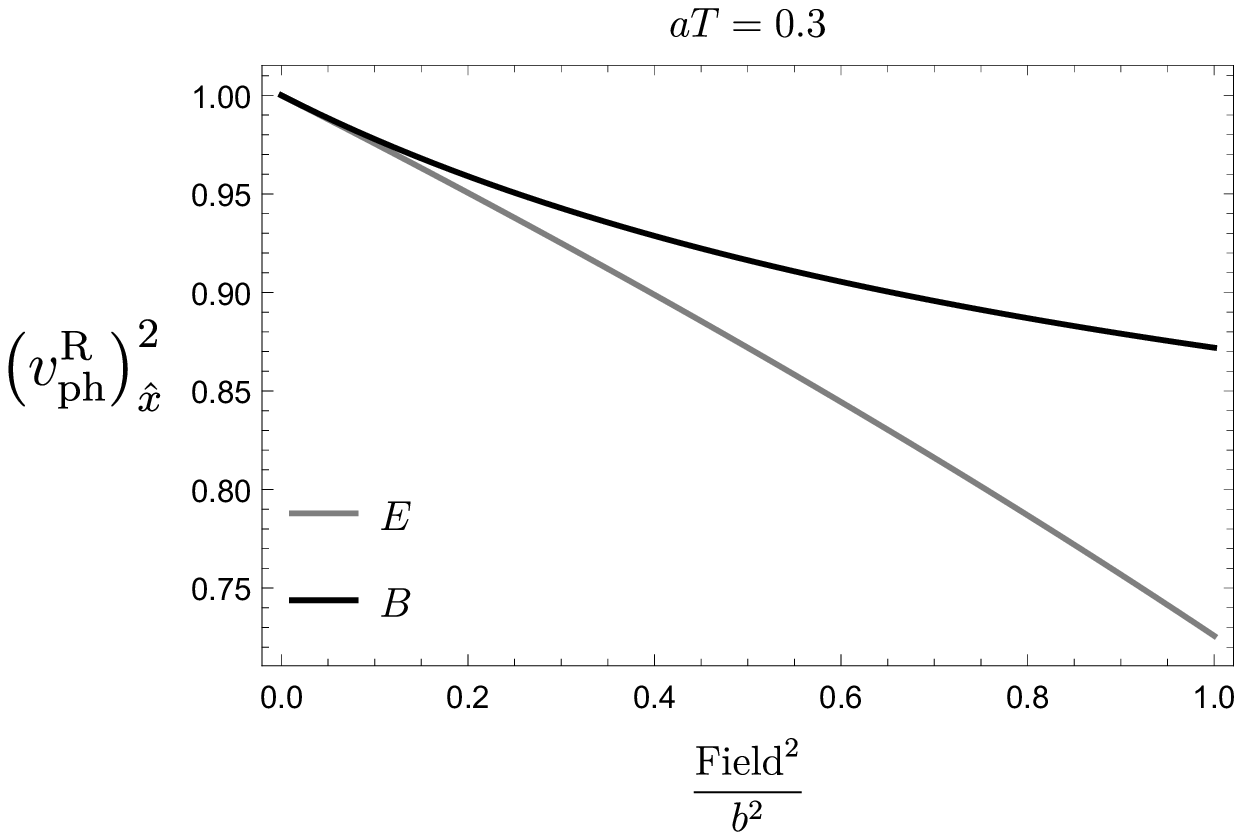}}
\caption{\small   The exact expressions, Eqs. (\ref{vRindEzp}) and (\ref{vRindExp}), are shown for fixed $aT$. The black curve is for a BI magnetic background and the gray corresponds to an electric background.
The upper plots correspond to the waves moving in the $\pm \hat{z}$ directions, and the bottom to the $\hat{x}$ direction. We consider that the components of the background fields have the same intensities, $B_{x}=B_{z}$, and  $E_{x}=E_{z}$. }
\label{f:Fig10}
\end{figure}

In Fig. \ref{f:Fig10} are plotted the exact expressions for the phase velocities,   Eqs. (\ref{vRindEzp}) and (\ref{vRindExp}), for the waves moving in  $\pm \hat{z}$ and $\hat{x}$ directions.

In $(a)$, for the waves moving in the $\pm\hat{z}$ direction, the lowest phase velocity is reached by the wave in an electric background. The wave moving in the $\hat{z}$ direction reaches lower values in both background fields. The behavior of the wave moving in the $\hat{x}$ direction is similar to the one in $-\hat{z}$ but it reaches slightly higher values in both background fields.  

Note in Figure \ref{f:Fig10} that for the range $\frac{{\text Field}^2}{b^2}< 0.2$
it is not possible to distinguish between the magnetic and electric background as analyzed previously.

\begin{figure}[H]
 \centering
\subfigure[]{ \includegraphics[width=0.45\textwidth]{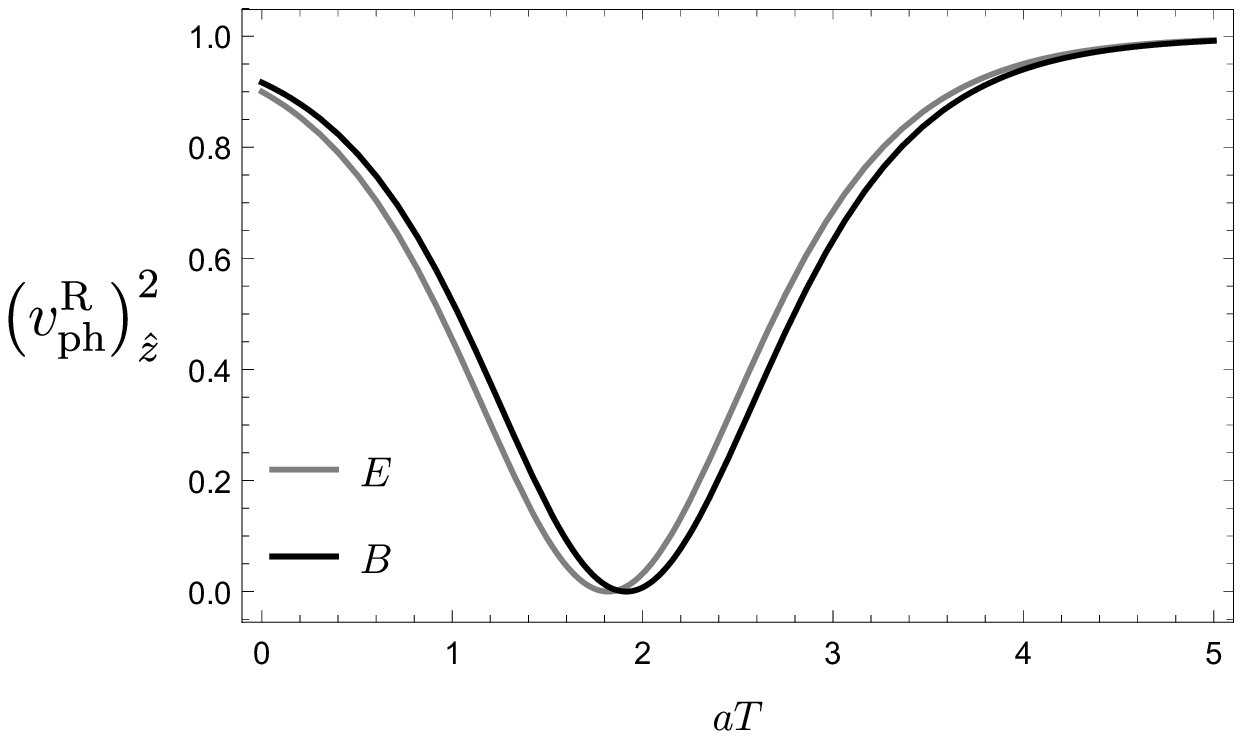}}
\subfigure[]{ \includegraphics[width=0.45\textwidth]{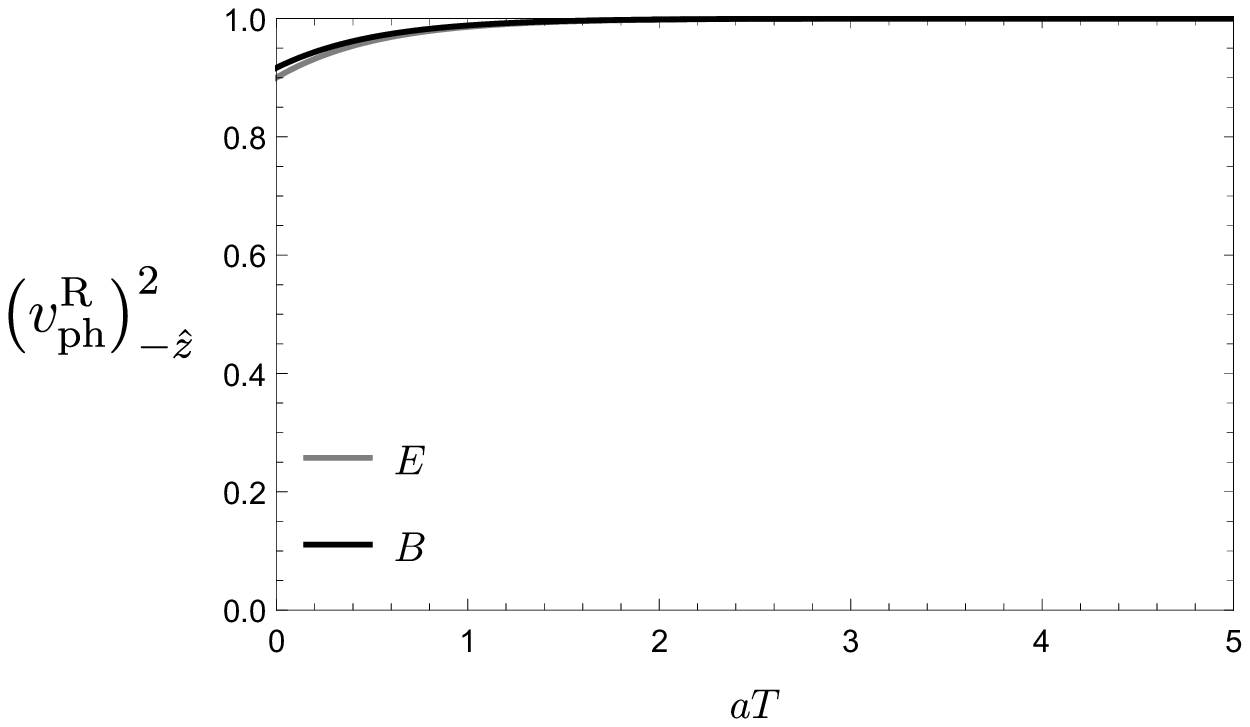}}
\subfigure[]{\includegraphics[width=0.45\textwidth]{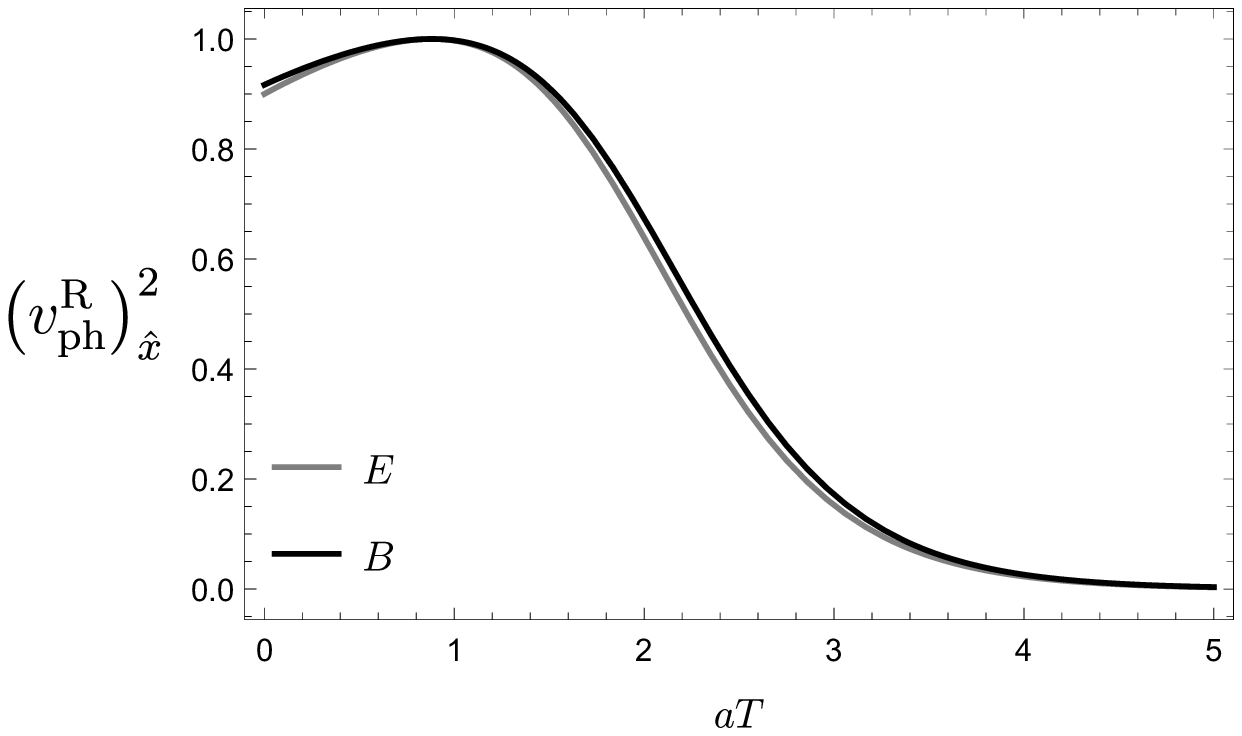}}
\caption{\small   The exact expressions, Eqs. (\ref{vRindEzp}) and (\ref{vRindExp}), are shown  as a function of $aT$, for a fixed BI background.
The components of the background fields are of the same magnitude, $B_{x}=B_{z}$, $E_{x}=E_{z}$, and  $B^2 /b^2 =0.2$, $E^2 /b^2 =0.2$; the behavior of the phase velocity is similar in both backgrounds.}
\label{f:Fig11}
\end{figure}

For completeness in Fig.  \ref{f:Fig11} we compare the behavior of the phase velocities as a function of $aT$ for fixed electric and magnetic BI background.
Comparing  $(a)$ and  $(b)$ note that the behavior is similar for both background fields.
For the waves moving in the $\pm \hat{z}$ directions, as $aT$ is higher the phase velocity tends to the light velocity in vacuum. The phase velocity tends to zero as $aT$ increase for the wave moving in $\hat{x}$ direction. 
\section{The Redshift of light pulses propagating through a BI background in the Rindler frame}

Gravitational redshift is the increment of the wavelength of electromagnetic radiation due to the presence of a gravitational field.
According to the EEP, an accelerated frame is equivalent to a gravitational field, therefore it is expected that light pulses sent from one Rindler observer to another will modify their frequency, resembling what happens in the presence of a gravitational field.
This is indeed what happens, see  \cite{Kooks2020}, \cite{Hamilton1978}, \cite{Aiello2008}, \cite{Alberci2018}.

In this section, we determine how the redshift due to the acceleration of the frame is affected when additionally there is a BI magnetic background.

\begin{figure}[H]
\centering
\includegraphics[width=0.6\textwidth]{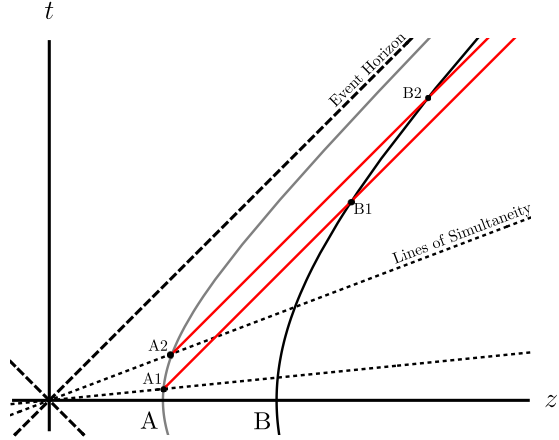}
\caption{\small  In this diagram are shown,  in the Lab frame coordinates $(z,t)$,  the world lines (hyperbolas) of  two Rindler observers, A and B;  the (dotted) lines of simultaneity of the two light rays emitted from $A$ to B as well as the trajectories of the two light pulses (in gray).  It is illustrated that the proper time intervals are different, $\Delta\tau_{r}=\tau_{B2}-\tau_{B1} >  \Delta\tau_{e}=\tau_{A2}-\tau_{A1}.$  }
\label{f:Fig12}
\end{figure}

\begin{figure}
    \centering
    \includegraphics[width=0.6\textwidth]{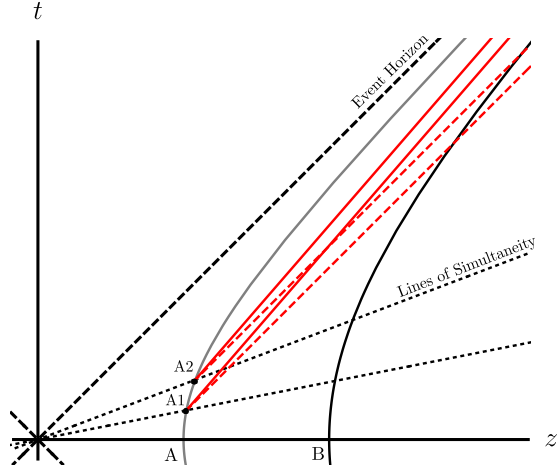}
    \caption{\small It is shown the difference between the light rays in vacuum (dashed gray) and the presence of a  magnetic BI background (continuous gray).
The difference between the intervals is larger in the presence of the BI field, which means a larger redshift than the one due only to the acceleration of the Rindler frame.  }
  \label{f:Fig13}
\end{figure}

The redshift, denoted as  $z_{R}$,  can be written  in terms of the intervals in proper time of emission ${\Delta \tau_{e}}$ and reception ${\Delta \tau_{r}}$ of two light rays with wavelength $\lambda$ and frequency $f$ as:
\begin{equation}
z_{R}+1=\frac{\lambda_{r}}{\lambda_{e}}=\frac{f_{e}}{f_{r}}=\frac{\Delta \tau_{r}}{\Delta \tau_{e}}> 1
\end{equation}
where the subscripts $e$ and $r$ refer to the emitter and receiver, respectively.
Therefore to determine the redshift in the Rindler frame we first have to calculate the proper time intervals elapsed between the two sent signals, ${\Delta \tau_{e}}$  and the interval elapsed between the reception of the same two pulses, ${\Delta \tau_{r}}$, as measured by the receiver. These intervals will be different if a redshift occurs;  it is illustrated in Fig. \ref{f:Fig12}. In Fig. \ref{f:Fig13} is illustrated as well the intervals in the world lines of the Rindler observers when additionally there is a BI magnetic background.
  
 Considering that $A$ sends a light pulse, the light trajectory goes from the event of emission A1 to the reception B1. When $B$ receives the signal, its velocity with respect to the laboratory frame is higher than the velocity of $A$ at the moment of emission. Analogously for the second signal, in such a way that the proper time intervals of reception and emission, respectively,  are given by, 

\begin{equation}
 \Delta\tau_{r}=\tau_{B2}-\tau_{B1}, \quad  \Delta\tau_{e}=\tau_{A2}-\tau_{A1}.
\end{equation}

The $(z,t)$ coordinates in the Lab frame in terms of the proper time $\tau_{i}$ and the acceleration $a_{i}$  of each observer are

\begin{equation}
z_i=\frac{1}{a_{i}}{\rm Ch}({a_{i}\tau_{i}}), \quad t_i=\frac{1}{a_{i}}{\rm Sh}({a_{i}\tau_{i}}), \quad i= A,B.   
\label{tz_coord} 
\end{equation}

As each observer is considered the principal observer, their proper positions are $Z_{i}=0$.

To determine the redshift we restrict ourselves to a wave moving in the $(+ \hat{z})$ direction in a magnetic BI background located in the plane XZ and the Rindler acceleration being $\vec{a}= a \hat{z}$. The light  trajectory, in Minkowski coordinates $(t,z)$,  is calculated integrating the phase velocity in Eq. (\ref{velAiello}),  $v_{\rm ph} = \beta=\frac{dz}{dt}=\sqrt{ 1-\frac{B_{x}^2}{b^2+B^2}}$,

\begin{equation}
 z-z_{0}= \sqrt{ 1-\frac{B_{x}^2}{b^2+B^2}}(t-t_{0})=\beta(t - t_{0}),
\end{equation}
where   the phase velocity is denoted by $\beta$, 
when $\beta\rightarrow 1$ (zero BI field)  it is recovered  the trajectory of   light in vacuum, this case is treated in \cite{Kooks2020}.

Considering that
the first light pulse  is emitted by A at the initial coordinates: $z_{0}=z_{1}$ and $t_{0}=t_{1}$, then the trajectory of the light pulse is $z-z_1= \beta (t-t_1)$,  that in Rindler coordinates,  Eqs. (\ref{tz_coord}), amounts to 

\begin{equation}
z-\frac{1}{a_{A}} {\rm Ch}{(a_{A}\tau_{A1})}=\beta [t-\frac{1}{a_{A}}{\rm Sh}(a_{A}\tau_{A1})];
\end{equation}
this light ray intersects the world line of $B$  at  $(z_{3}, t_{3})$ when the proper time of $B$ is $\tau_{B1}$. Since $t_{3}= {\rm Sh} (a_{B} \tau_{B1})/a_{B}$, then knowing $t_{3}$ we can determine $\tau_{B1}$. Solving for $t_{3}$ implies to solve the system of equations consisting on the  hyperbola equation for B and the light trajectory equation, i.e. solving the system

\begin{eqnarray}
 z_{3}^2&=& \frac{1}{a_{B}^2}+t_{3}^2 \\
z_{3}-z_{1} &=&  \beta (t_{3}-t_{1}), 
\end{eqnarray}
from these equations it is obtained a quadratic equation for $t_{3}$ with solution

\begin{equation}
a_{A}a_{B} t_{3}=\frac{1}{(\beta^2-1)} \left[ \beta a_{B} \chi_{1} \pm \sqrt{a_{B}^2 \chi_{1}^2 + a_{A}^2 (\beta^2-1)} \right],
\end{equation}
where $\chi_{i}= a_{A}(\beta t_{i}-z_{i})$
that in terms of the Rindler coordinates for the emitter A, is

\begin{equation}
\chi_{i}=a_{A} (\beta t_{i}-z_{i}) = \left[{ \beta \rm Sh} (a_{A}\tau_{A1}) - {\rm Ch } (a_{A}\tau_{A1}) \right].
\end{equation}
Using $t_{3}$ in Rindler coordinates, $t_{3}={\rm Sh}{(a_{B}\tau_{B1})}/a_{B}$, we obtain

\begin{eqnarray}
 {\rm Sh}{(a_{B}\tau_{B1})} & = &\frac{1}{a_{A}(\beta^2-1)} \left(\beta a_{B} \chi_{1} \pm \sqrt{a_{B}^2 \chi_{1}^2+a_{A}^2 (\beta^2-1)} \right), \rightarrow  \nonumber\\
a_{B}\tau_{B1} & = & {\rm ArcSh} \left[ \frac{1}{a_{A}(\beta^2-1)} \left(\beta a_{B} \chi_{1} \pm \sqrt{a_{B}^2 \chi_{1}^2+a_{A}^2 (\beta^2-1)} \right) \right] \nonumber\\
 & = & {\rm ArcSh}( g ), 
\label{tauB1}
\end{eqnarray}

Following the same procedure for the second light ray, we obtain: 

\begin{eqnarray}
{\rm Sh}{(a_{B}\tau_{B2})} & = & \frac{1}{a_{A}(\beta^2-1)} \left(\beta a_{B} \chi_{2} \pm \sqrt{a_{B}^2 \chi_{2}^{2}+a_{A}^2 (\beta^2-1)} \right), \rightarrow 
\nonumber\\
 a_{B}\tau_{B2} & = &  {\rm ArcSh}\left[\frac{1}{a_{A}(\beta^2-1)} \left(\beta a_{B} \chi_{1} \pm \sqrt{a_{B}^2 \chi_{1}^2+a_{A}^2 (\beta^2-1)} \right) \right] \nonumber\\
 & = & {\rm ArcSh}( f ),  
\label{tauB2}
\end{eqnarray}
where we have defined $f$ and $g$ as the right hand sides of the previous equations.
Once determining $\Delta\tau_{r}=\tau_{B2}-\tau_{B1} $ and $\Delta\tau_{e}=\tau_{A1}-\tau_{A2}$ from the previous expressions  we can measure the redshift of the two light pulses propagating through  a  BI magnetic background  in the Rindler frame.

From (\ref{tauB2}) and (\ref{tauB1}) the proper time interval of the reception of the pulses is,

\begin{equation}
 a_{B}(\tau_{B2}-\tau_{B1})={\rm ArcSh}{f}-{\rm ArcSh}{g}=\log{\frac{\sqrt{f^2+1}+f}{\sqrt{g^2+1}+g}};
\end{equation}
expanding the result for "small" intensities of the field, i.e. neglecting terms of order $(B/b)^4$ and higher,

 \begin{equation}
    \tau_{B2}-\tau_{B1}\approx \frac{a_{A}}{a_{B}} (\tau_{A2}-\tau_{A1}) + \frac{ \left(\left(a_A^2-a_B^2\right) \left(e^{2 a_A \tau _{\text{A2}}}-e^{2 a_A \tau _{\text{A1}}}\right)\right)}{4 a_B^2}\frac{B_{x}^2}{b^2},
\end{equation}
 in this equation the first term corresponds  to the redshift  due to the acceleration of the observer´s frame,  while the second term is the contribution due to the presence of the BI electromagnetic background,  that   clearly depends on the magnetic BI component that is transversal to the acceleration of the Rindler frame.   We can obtain a simpler expression approximating for small proper times; in this case we approximate the exponentials as $e^{x} \approx 1+x$, neglecting  terms  $(a_{A} \tau_{A})^2$  and higher,  

\begin{equation}
    \tau_{B2}-\tau_{B1} \approx \frac{a_{A}}{a_{B}}\left(\tau_{A2}-\tau_{A1} \right) \left(1+\frac{a_{A}^2-a_{B}^2}{2a_{B}^2}\frac{B_{x}^2}{b^2} \right),
\end{equation}

\begin{equation}
   \Delta \tau_{B} \approx \frac{a_{A}}{a_{B}}\Delta\tau_{A} \left(1+\frac{a_{A}^2-a_{B}^2}{2a_{B}^2}\frac{B_{x}^2}{b^2} \right).
   \label{redshift1}
\end{equation}
To obtain  (\ref{redshift1}) in terms of the frequency we note that if $A$ sends a pulse with a proper frequency $f_{A}$, this means that $A$ measures the number of waves per unit of his proper time. While $B$ receives and measures a proper frequency $f_{B}$.  Since the  number of light pulses is the same,  

\begin{equation}
    f_{A}\Delta \tau_{A}=f_{B} \Delta \tau_{B},
\end{equation}
then
\begin{equation}
   \frac{f_{A}}{f_{B}}=\frac{\Delta \tau_{B}}{\Delta\tau_{A}} \approx \frac{a_{A}}{a_{B}} \left(1+\frac{a_{A}^2-a_{B}^2}{2a_{B}^2}\frac{B_{x}^2}{b^2} \right),
  \end{equation}
Since  the proper accelerations are $a_{A}/ a_{B}> 1$, and the BI term is always positive, then for the frequencies it is true that $f_{A}/f_{B}>1$, and
in terms of the redshift parameter $z_{R}=\frac{f_{A}}{f_{B}}-1$

\begin{equation}
    z_{R}=\frac{a_{A}}{a_{B}} \left(1+\frac{a_{A}^2-a_{B}^2}{2a_{B}^2}\frac{B_{x}^2}{b^2} \right)-1.
    \label{zRredshift}
\end{equation}

\begin{figure}[H]
\centering
\includegraphics[width=0.6\textwidth]{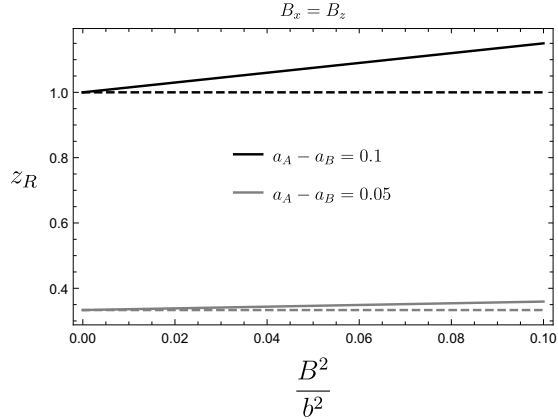}
\caption{\small It is plotted the redshift in Eq.  (\ref{zRredshift}) with respect to the intensity of the BI magnetic background, for two different
proper acceleration differences between the emitter (A) and the receiver (B);  as the difference becomes smaller, the resulting redshift is smaller as well. 
 The dashed lines represent the pure Rindler effect, i.e. the one corresponding to a vanishing BI magnetic background, while the continuous
lines account for the total redshift; the difference between the two lines (continuous minus dashed) corresponds to the BI redshift.
In the plot the magnetic components are of the same magnitude, $B_{x}=B_{z}$. }
\label{f:Fig14}
\end{figure}

Analyzing the redshift expression we see that there is a loss of energy by the light pulse; this loss is composed of two terms, a first part is spent in overcoming the gravitational field (Rindler acceleration), as expected;  plus the effect of the BI field, i.e. the pulse has to spend energy as well while traveling through the magnetic background, resulting in a larger redshift. In Fig. \ref{f:Fig14} are illustrated two examples.

Another way of writing the redshift is in terms of the position of the observers. 
Considering the coordinate transformation to $(\bar{T}, \bar{Z})$ coordinates
\begin{equation}
    \bar{Z}=Z+\frac{1}{a}, \qquad \bar{T}=T, \qquad \bar{X}=X,\qquad \bar{Y}=Y;
\end{equation}
then
\begin{equation}
   \left\{\begin{array}{c}
        t \\
        z 
    \end{array}
    \right\}=\bar{Z}\left\{
    \begin{array}{c}
    {\rm Sh} a \bar{T} \\ 
    {\rm Ch} a \bar{T}
    \end{array} \right\}\rightarrow
    \begin{array}{c}
    \bar{Z}=\sqrt{z^2-t^2}\\
    \bar{T}=\frac{1}{a}{\rm ArcTh}(\frac{t}{z});
    \end{array}
\end{equation}
while the Rindler metric takes the form $ds^2=a^2\bar{Z}^2d\bar{T}^2-d\bar{Z}^2-d\bar{X}^2-d\bar{Y}^2$; in these coordinates the event horizon is at 
$\bar{Z}=0$. 

\begin{figure}[H]
\centering
\includegraphics[width=0.6\textwidth]{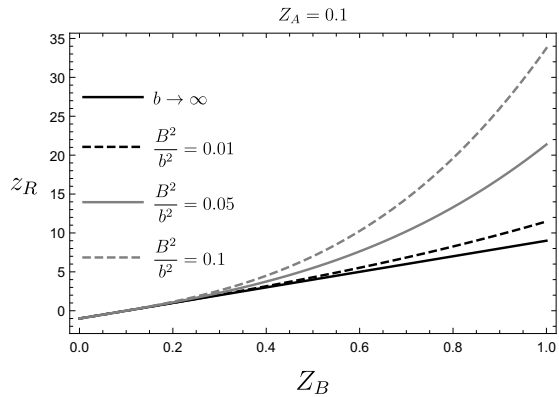}
\caption{\small It is plotted the redshift in Eq.  (\ref{zRredshift2}) with respect to the proper coordinates of the receiver ${Z}_{B}$, for different values of the intensity of the BI field;  the redshift increases as the intensity of the field increases. In this plot $Z_{a}=0.1$. As farther is the receiver from the emitter the redshift is larger. }
\label{f:Fig15}
\end{figure}

To calculate the redshift we write the proper coordinates of the emitter (A) and the receiver (B) as: 
\begin{equation}
    A:\{\bar{\tau}_{A}, \bar{Z}_{A}=\frac{1}{a_{A}} \}, \qquad B: \{\bar{\tau}_{B},\bar{Z}_{B}=\frac{1}{a_{B}}\}
\end{equation}
Note that now the proper position of the observers is at $\bar{Z}_{i}={1}/{a_{i}}$. The Minkowski coordinates corresponding to the positions of emission and reception of the light pulses are Eqs. (\ref{tz_coord}), then we determine  the redshift analogously and Eq. ( \ref{zRredshift}) can be written in terms of $(\bar{T}, \bar{Z})$ as 

\begin{equation}
    z_{R}=\frac{\bar{Z}_{B}}{\bar{Z}_{A}}\left(1+\frac{\bar{Z}_{B}}{\bar{Z}_{A}}\frac{B_{x}^2}{2b^2} \right)-1.
    \label{zRredshift2}
\end{equation}
In Fig. \ref{f:Fig15} is shown the redshift as a function of the receiver position, for a fixed emitter, for different field intensities.

\section{Conclusions}

We have considered an electromagnetic wave,  or light ray, propagating through an intense uniform Born-Infeld (BI) background, and have determined the phase velocities measured by an accelerated observer. 
 For the accelerated frame we have considered a Rindler spacetime. This situation models as well an environment with a uniform gravitational field, according to the Einstein Equivalence Principle.
The phase velocities are determined from the effective optical metric, that is a curved spacetime produced by the presence of the intense BI magnetic or electric field.

Using the NLED  effective optical metric approach \cite{Novello2000}, \cite{Pleban} and then applying a Rindler transformation to it   we obtain 
the phase velocity of the propagating wave from the null geodesics of the transformed effective optical metric.
Our treatment is valid for very strong fields; if we consider, for instance, that $B^2/b^2 \approx 10^{-2}$, then the BI field is of the order of $10^{19}$V/m that is ten times the critical Schwinger field or  $B_{\rm cr} \approx 10^{9}$ Tesla. 
We have considered first a uniform BI magnetic background and then a  purely electric one and three different directions of the propagating wave; the setting is shown in Fig. \ref{f:Fig3}.

For the BI  magnetic background,  the phase velocity of the propagating light significantly slows down for a wave moving in the same direction as the Rindler acceleration, diminishing as $B$ grows;  on the contrary, the phase velocity goes faster for a wave moving in the directions that are opposite and transversal to the Rindler acceleration, that we have considered being $a \hat{z}$.
The phase velocities depend on $(aT)$ such that the effect of increasing the acceleration is the same as the one of time elapsing.
If the magnetic field component transversal to the acceleration vanishes, then there is no effect on the phase velocity neither of the acceleration nor the BI field, and its value is the one in vacuum. 

For fixed values of the BI magnetic field phase velocity of the waves moving in  $+ \hat{z}$  and $+ \hat{x}$ directions the decreases reaching zero and then increases again as $aT$ increases; while the phase velocity of the wave propagating in the $-z$ direction approaches the one in vacuum as $aT$ increases. In Figs. \ref{f:Fig7} and \ref{f:Fig8}  these behaviors are shown for the squared phase velocities.
Recall that these are the velocities as measured by the accelerated observer,  and the relative directions change as $aT$ increases:  initially, the observer chases the light ray, and as it approaches the wave, the wave's velocity seems to decrease, and eventually, it is zero (the moment the observer reaches the wave) and then the relative direction changes, since subsequently, the observer moves away from the wave, then wave's velocity starts to increase.  Since the accelerated frame is no longer inertial, no special relativity velocity invariance is expected.
We also address the situation of a propagating wave through
an intense electric background field. 
The effect of the BI electric field (slowing down the phase velocity) is qualitatively very similar to the one of the BI magnetic background, and, in the approximation taken up to $B^2/b^2$ terms, the expressions for the phase velocities are the same just changing $B_{i} \mapsto E_{i}$.  For $E_{z}=0$ the slowing down is maximized. The wave traveling in $- \hat{z}$ is not affected by the BI field and as $aT$ increases the phase velocity reaches the one in vacuum.
If the electric field component that is perpendicular to the acceleration vanishes, then there is no effect neither of the acceleration or the BI field on the phase velocity, and its value is the one in vacuum. 
For strong fields, when $B$ approaches the maximum attainable BI field $b$, the behavior of the phase velocities is quantitatively different depending on if the background is electric or magnetic; the most effective for slowing down the phase velocity is the electric background, for all directions of the light ray.

Finally, we analyzed the redshift of a light pulse sent from one Rindler observer and received by another when the light pulses travel through the BI magnetic background. From the trajectory of the pulse and the hyperbola worldline of the emitter and receiver, we determined the proper time intervals elapsed between the emission of the two pulses and then the proper time interval of the reception. 
Using these intervals we calculated the redshift; then in the approximation of small fields  (that as we commented above, are still very intense fields) and small intervals of time, we find the expression for the redshift: two contributions can be distinguished, one due to the acceleration of the frame and an additional redshift is produced by the presence of the BI magnetic background, resulting then in a larger redshift.

In summary, we have analyzed the phase velocity of light propagating under the effect of a very intense magnetic or electric BI field as measured by an accelerated (Rindler) observer;  according to the EEP, the situation is equivalent to the measurements by an observer under the influence of a uniform gravitational field.

\vspace{0.5cm}
\textbf{Acknowledgments}: The work of E G-H has been sponsored by CONACYT-Mexico through the Ph. D.  scholarship No.761878.  N. B. acknowledges partial financial support from CONACYT-Mexico through project No. 284489.

\end{document}